\documentclass[a4paper,11pt]{article}
\usepackage{jcappub} % for details on the use of the package, please see the JINST-author-manual
\usepackage{natbib}
\usepackage{multirow}
\usepackage{float}
\usepackage{comment}
\usepackage{cancel}
\usepackage[normalem]{ulem} 
\usepackage{soul}
\usepackage{siunitx}
\usepackage{graphicx}
\usepackage{gensymb}
\usepackage{amsmath, cases}
\usepackage[usenames]{color}
\usepackage{amssymb}
\usepackage{hyperref}
\usepackage{cleveref}
\newcommand{\be}{\begin{equation}}
\newcommand{\ee}{\end{equation}}
\newcommand{\bea}{\begin{aligned}}
\newcommand{\eea}{\end{aligned}}
\newcommand{\pr}{\partial}

\newcommand{\bse}{\begin{subequations}}
\newcommand{\ese}{\end{subequations}}

\newcommand{\bmm}{\begin{multline}}
\newcommand{\emm}{\end{multline}}

\begin{document}
\title{Resonant transmission of scalar waves through rotating traversable wormhole}
\author[a]{Rajesh Karmakar}
\emailAdd{rajesh@shu.edu.cn}
\affiliation[a, b]{Department of Physics, Shanghai University, 99 Shangda Road, Shanghai, 200444, China}
\author[b]{Bum-Hoon Lee}
\emailAdd{bhl@sogang.ac.kr}
\affiliation[b, c]{Center for Quantum Spacetime, Sogang University, Seoul 04107, Korea
}
\affiliation[b]{Department of Physics, Sogang University, Seoul 04107, Korea
}
%\affiliation[b]{Department of Physics, Shanghai University, 99 Shangda Road, Shanghai, 200444, China}
\author[c]{Wonwoo Lee}
\emailAdd{warrior@sogang.ac.kr}
%\affiliation[c]{Center for Quantum Spacetime, Sogang University, Seoul 04107, Korea
%}

\abstract{The viability of traversable wormholes as exotic compact objects relies on identifying signatures that distinguish them from black holes and other compact objects. In this work, we investigate the propagation of a massless scalar field in a rotating traversable wormhole described by Teo’s class of solutions. We compute the greybody factor (transmissivity) and identify sharp Breit–Wigner-type resonances associated with scalar modes trapped in the potential well surrounding the throat. Rotation modifies these resonances through frame dragging, with counter rotating modes developing deeper trapping wells. We further investigate the stability of the long lived resonant modes through their quasinormal mode spectrum and find that corotating modes exhibit ergoregion instability. We therefore focus on the partial transmission cross section of the stable counter rotating modes, which characterizes the resonant transmission of incident scalar flux through the wormhole to the other asymptotic region. Our findings establish the role of rotation and mode stability in shaping the resonant transmission of scalar waves through traversable wormholes, providing a basis for future studies of analogous resonant phenomena in electromagnetic and gravitational waves, and their potential observational implications.}

\maketitle

\section{Introduction}
The Universe is undergoing dynamic evolution, with its primary energy budget consisting of time-evolving dark energy and dark matter. Consequently, investigations of solutions that coexist with matter fields, along with vacuum solutions to the Einstein equations, 
will continue to be of profound interest. Over the past decade, there has been significant progress in probing the strong field regime of gravity through gravitational wave detections and observations of black hole (BH) shadow images. However, these measurements are primarily sensitive up to the photon sphere, leaving open the possibility that exotic compact objects may exist alongside vacuum BH solutions \cite{LIGOScientific:2021sio, Vagnozzi:2022moj}.   

Wormholes (WHs) are extraordinary solutions that go beyond the vacuum solutions allowed by general relativity. Einstein and Rosen were the first to conduct a serious study of WH physics, considering a bridge connecting two identical sheets of spacetime~\cite{Einstein:1935tc}. This object has been extensively studied alongside its various physical properties~\cite{Misner:1957mt, Ellis:1973yv, Chetouani:1984qdm} and has undergone a modern revitalization with the work of Morris and Thorne~\cite{Morris:1988cz}. Their proposal initiated the traversable scenario through a WH throat, in a more or less geometric setting. However, it has been pointed out that traversability requires exotic matter to prevent the throat from collapsing. In contrast to the usual gravitating matter, this leads to the violation of the null energy condition~\cite{Lobo:2005us, Lobo:2020xvs, Churilova:2021tgn, Antoniou:2019awm}. Even before the Morris-Thorne proposal, a phantom-like scalar field with negative kinetic energy was shown to support a traversable WH \cite{Ellis:1973yv}. Subsequently, extensions involving charged scalar fields have been developed, resulting in charged WHs~\cite{Huang:2019arj, Godani:2020gbr, Koga:2025bqw, Rahaman:2025jhv, Turimov:2025zxy, Blazquez-Salcedo:2025dit}. The concept of `charge without charge' has been introduced in Ref.~\cite{Misner:1957mt} and its geometric realization has been explored~\cite{Kim:2001ri, Kim:2024mam, Kim:2025zyo, Dzhunushaliev:2025ngw}. 
Meanwhile, traversable WHs and their properties have been extensively investigated in a wide range of alternative theories of gravitation~\cite{Kim:1997jf,  Lobo:2009ip, Kanti:2011jz, Kim:2016pky, Ovgun:2018xys, Halder:2019urh, Kim:2019ojs, Hassan:2021egb, Bouhmadi-Lopez:2021zwt, DeFalco:2023twb, Karmakar:2026yqg}.   
From a geometrical standpoint, alternative approaches have sought to construct traversable WH geometry following the cut-and-paste mechanism. Initial works in this direction involve the thin shell WH by Visser \cite{Visser:1995cc}.
Although a large amount of literature is devoted to the investigation of spherically symmetric spacetimes (see \cite{Visser:1995cc} for an extensive discussion), a simultaneous effort has also led to the development of axisymmetric and generically asymmetric WH geometries \cite{Hoffmann:2018oml, Forghani:2018gza, Forghani:2018fks}.

Of particular importance in the present work are rotating axisymmetric WHs, which carry several motivations. Astrophysically, most compact objects are rotating, which makes the inclusion of angular momentum essential for realistic models. In fact, rotation has been shown to improve the stability of WH spacetimes \cite{Azad:2023iju}. In contrast to the static case, the presence of rotation sometimes also reduces the violation of the null energy condition \cite{Kashargin:2008pk, Uemichi:2026dzb}. From an observational perspective, rotating WHs may support accretion disks, whose properties could differ from those around rotating BHs, providing potential signatures to distinguish between the two \cite{Harko:2009xf, Karimov:2019qfw, Paul:2019trt}. 

Without explicit construction, Teo proposed a rotating scenario, extending the Morris-Thorne-type traversable WH, completely based on the symmetry arguments \cite{Teo:1998dp}. Later, the possibility of a source term for such a rotating WH spacetime has been studied~ \cite{PerezBergliaffa:2000ms}. Afterwards, several axisymmetric WHs have been proposed \cite{kashargin2008slowly, Mazza:2021rgq, Kashargin:2008pk, Volkov:2021blw, Cisterna:2023uqf}. Although observable signatures of such WH geometries have not yet been detected, a range of scenarios regarding their existence have been explored, along with studies of their potential astrophysical implications \cite{DeFalco:2023kqy, DeSimone:2025sgu, Tsukamoto:2012xs, Bambi:2013nla, Azreg-Ainou:2014dwa, Abdujabbarov:2016efm, Dai:2019mse, DeFalco:2020afv, Simonetti:2020ivl, Godani:2021aub, Sarkar:2024bxk}.

In this work, we investigate the propagation of a minimally coupled, massless scalar field in the background of a rotating traversable WH described by the Teo metric~\cite{Teo:1998dp}. Among the various rotating WH geometries \cite{Mazza:2021rgq, Volkov:2021blw, Cisterna:2023uqf}, the Teo WH provides a relatively simple framework for investigating wave propagation, particularly because its frame-dragging angular velocity has a simple radial dependence, allowing the influence of rotation on different field modes to be studied more directly. Nevertheless, we adopt the test-field approximation, neglecting the backreaction of the scalar field on the background geometry. The scattering and transmission properties of WH spacetimes, particularly in the static case, have been investigated in previous studies, revealing distinctive features compared to those of BHs~\cite{Kar:1995jz, Lima:2020auu, LimaJunior:2022zvu}. In particular, scalar modes temporarily trapped within the effective potential well surrounding the throat can give rise to sharp resonant peaks in the transmission spectrum. Unlike rotating BHs, where bosonic waves can undergo superradiant amplification in the low-frequency regime, rotating horizonless compact objects may instead exhibit an ergoregion instability \cite{Cardoso:2007az, Franzin:2022iai}, characterized by exponentially growing perturbation modes. Although these phenomena are closely related to the extraction of rotational energy for BHs, they manifest differently due to the absence of an event horizon in the latter case \cite{Friedman:1978wla, Maggio:2017ivp, Siemonsen:2025wib}. The absence of superradiant amplification in the scattering of scalar waves by rotating WHs was first pointed out in Ref.~\cite{Konoplya:2010kv}, where the reflection and transmission coefficients were also computed in the static case. Hence, it is imperative to systematically investigate the influence of rotation on resonant transmission and the stability of the associated modes, particularly in the low-frequency regime, where the scattering properties of rotating BHs and WHs may differ significantly.

Motivated by these considerations, we numerically compute the reflection and transmission coefficients to determine the greybody factor (transmissivity) and investigate the resonant characteristics of scalar waves propagating through the rotating Teo WH~\cite{Teo:1998dp}. To the best of our knowledge, a systematic investigation of these resonant transmission features in this rotating WH background has not previously been carried out. We further analyze the quasinormal mode (QNM) spectrum, focusing particularly on the long lived modes responsible for the resonant peaks and assessing their stability in the presence of an ergoregion. Our analysis reveals that corotating modes can exhibit ergoregion instability, whereas the counter rotating modes remain stable while displaying pronounced resonant transmission. These findings are consistent with the earlier study of Kerr-like WHs \cite{Mazza:2021rgq} in Ref.~\cite{Franzin:2022iai}, which primarily investigated their QNM spectrum and stability properties. In contrast, the present work focuses on resonant scalar wave transmission and its connection to the QNM spectrum, while also considering the implications of mode instability for the scattering properties. Additionally, we define and analyze the partial transmission cross sections of the stable counter-rotating modes, which characterize the contributions of individual modes to the incident scalar field flux transmitted through the throat into the other asymptotic region.

%Correspondence of superposed QNMs and greybody factor ...late time echoes in the gravitational spectrum..observational uncertainties.Ergoregion and frame dragging...  

The remainder of the paper is organized as follows. In \Cref{sec: Teo_spacetime}, we briefly discuss the rotating traversable WH spacetime as proposed by Teo. In \Cref{sec: scalar_dynamics}, the governing equations have been derived for a minimally coupled massless scalar field. In \Cref{sec: gbody}, we explain the absence of superradiant amplification and present the definition and numerical analysis of the greybody factor. We also compute the scalar field QNM spectrum and discuss the stability properties of the corresponding modes, including the emergence of unstable modes. In \Cref{sec: norm_planewave}, we derive the normalization factor for scalar waves incident at an arbitrary angle relative to the rotation axis. Subsequently, in \Cref{sec: defn_pcs}, we define the partial transmission cross sections and discuss the distinctive features of the resulting spectra. Finally, in \Cref{sec: concl}, we conclude with the future outlook of the present work.  
%%%%%%%%%%%%%%%%%%%%%%%%%%%%%%%%%%%%%%%%%%%%%%%%%%%%%%%%%%%%%%%%%%%%%%%%%%%%%%%%%
\section{Brief Discussion of Teo-Class wormholes}\label{sec: Teo_spacetime}
The class of WH solutions introduced by Teo \cite{Teo:1998dp} constitutes a generalization of the Morris–Thorne traversable WHs, extending them from a static spherically symmetric configuration to a stationary axially symmetric framework. The associated spacetime metric, in its original form as proposed by Teo, is expressed as
\be\label{eq: Teo_metric}
\bea
ds^2&=-N^2dt^2+\frac{dr^2}{\left(1-\frac{b}{r}\right)}+r^2K^2\left[d\theta^2+\sin^2\theta(d\varphi-\Omega dt)^2\right],
\eea
\ee
where the four functions $N, b, K$, and $\Omega$  generally depend on $(r,\theta)$. The geometry of the WH is encoded primarily in the shape function $b(r, \theta)$, which sets the position of the throat.  From the shape function written in the Morris-Thorne form \cite{Morris:1988cz}, the position of the throat can be easily identified by the {\it apparent singularity} at $r=b$. However, the lapse function $N$, representing the gravitational redshift, must remain finite and nonzero throughout the domain to ensure traversability. In the angular part, the metric function $K$ is regular, positive, and non-decreasing, in the sense that $\pr_r R>0$, with $R=rK$ denoting the proper radial distance. The rotation is governed by the function $\Omega$, which represents the angular velocity of the WH. Furthermore, to ensure that the metric is non-singular on the rotation axis, the derivatives of $N$, $b$ and $K$ with respect to $\theta$ should vanish at $\theta = 0$ and $\theta = \pi$. 

For the WH to be stable and possess a traversable throat, a key geometric constraint arises from the requirement that the WH open outward at the throat. This geometry can be further analyzed by embedding an equatorial spatial slice of the WH into a Euclidean space \cite{Morris:1988cz}. This leads to the essential flare-out condition \cite{Kim:2013tsa}, which can be quantified as $\pr_r b< 1$ at the WH throat. When this condition is met, the metric describes two symmetric regions joined by a traversable throat at $r=b=r_0$ and  $r_0\leq r< \infty$. Additionally, demanding asymptotic flatness requires, 
\be
\bea
&N(r)=1-\frac{M}{r}+\mathcal{O}\left(\frac{1}{r^2}\right),~~~\frac{b}{r}=\mathcal{O}\left(\frac{1}{r}\right),\\
&K=1+\mathcal{O}\left(\frac{1}{r}\right),~~~\Omega=\frac{2J}{r^3}+\mathcal{O}\left(\frac{1}{r^4}\right).
\eea
\ee
From this, the conserved charges of the Teo WHs can be identified as $M$ and $J$, corresponding to the ADM mass and angular momentum, respectively. Nevertheless, the existence of a stable and traversable throat necessarily entails the violation of the null energy condition (NEC). While this requirement was originally established for asymptotically flat, static, and spherically symmetric WHs \cite{Morris:1988cz}, it has since been shown to persist irrespective of asymptotic or global properties, as well as in more general, non-symmetric configurations \cite{Hochberg:1998ii}. For the present case of Teo WHs, the constraints imposed on the metric functions, as outlined so far, are sufficient to ensure that the NEC is violated near the throat \cite{Teo:1998dp}.

Maintaining the essential characteristics of the Teo WH spacetime, and motivated by phenomenological considerations \cite{Nedkova:2013msa, Gyulchev:2018fmd, Kumar:2023wfp}, we adopt the following parameterization of the metric functions:
\be\label{eq: metric_functions}
b(r)=r_0,~~N=e^{-r_0/r},~~\Omega=\frac{2J}{r^3},~~~K=1,
\ee
where the constant $r_0>0$ denotes the throat radius. This choice satisfies the flare-out condition~\cite{Morris:1988cz, Hochberg:1997wp, Kim:2013tsa}. The angular momentum $J$ is related to the dimensionless spin parameter $a$ through $a\equiv J/M^2$, with $M=r_0$ adopted throughout this work.\footnote{The dimensional rotation parameter $\textbf{a}$ in the Kerr BH geometry is related to the dimensionless spin parameter by $\textbf{a}=aM$.} The static traversable WH is recovered for $a=0$, whereas a finite rotation can generate an ergoregion, defined by $g_{tt}>0$, where the stationary Killing vector becomes spacelike. For the line element \eqref{eq: Teo_metric} with the metric functions in \eqref{eq: metric_functions}, 
\be
g_{tt}=-e^{-2r_0/r}+\frac{4J^2}{r^4}\sin^2\theta,
\ee
which reduces to $g_{tt}(r_0,\pi/2)=-e^{-2}+4a^2$ at the equatorial throat. An ergoregion therefore exists for $|a|>1/(2e)\simeq0.184$, extending outward from the throat around the equatorial plane, with its boundary determined by $g_{tt}=0$ \cite{Teo:1998dp}. Having specified the background geometry, we now investigate the propagation of a test scalar field and the influence of rotation on its dynamics.
%%%%%%%%%%%%%%%%%%%%%%%%%%%%%%%%%%%%%%%%%%%%%%%%%%%%%%%%%%%%%%%%%%%%%%%%%%
\section{A test scalar field minimally coupled with the wormhole spacetime}\label{sec: scalar_dynamics} 
On the background of the rotating traversable WH spacetime, outlined in the previous section, we consider a minimally coupled massless test scalar field, described by the following action: 
\be
\mathcal{A}=-\frac{1}{2}\int\sqrt{-g}d^4x\, \pr_\mu\phi\pr^\mu\phi,
\ee
where $g$ represents the determinant of the spacetime metric \eqref{eq: Teo_metric}, explicitly \eqref{eq: metric_functions}, 
\be
\sqrt{-g}=r^2N(r)\sin\theta/(1-r_0/r).
\ee
Given the above action, the equation of motion of the scalar field reads
\be
\pr_\mu\left(\sqrt{-g}g^{\mu\nu}\pr_\nu\phi\right)=0.
\ee
For the rotating traversable WH solution under consideration, the governing equation of the scalar field can be expressed as
\be
\bea
&-\frac{1}{N(r)^2}
\left(\pr_t+\Omega(r)\pr_\varphi \right)^2 \phi+\frac{\sqrt{1-\frac{r_0}{r}}}{r^2N(r)}\pr_r\left(
r^2N(r)\sqrt{1-\frac{r_0}{r}}\pr_r\phi
\right)\\
&+\frac{1}{r^2\sin^2\theta}\pr_\varphi^2 \phi
+\frac{1}{r^2\sin\theta}
\pr_\theta\left(\sin\theta\,\pr_\theta\phi\right)=0.\\
\eea
\ee
Let us decompose the field into spherical harmonics as
\be\label{eq: mode_decomposition}
\phi=\sum_{lm}\mathcal{N}_{\omega lm}e^{-i\omega t}R_{lm}(r)Y_{lm}(\theta, \varphi),
\ee
with $\mathcal{N}_{\omega lm}$ denoting the overall normalization constant, which will be fixed in later sections. Here, $\omega$ represents the frequency of the scalar field, while $l$ and $m$ represent the orbital angular momentum number and the magnetic number, respectively. The modes with $m > 0$ are corotating with the WH, while those with $m < 0$ are counter-rotating. Notably, for rotating BH geometry \cite{Macedo:2013afa}, one usually get the decoupled angular equation to result in spheroidal harmonics. In the special class of Teo WH, we find that spherical harmonics suitably decouple the equation of motion. Nevertheless, with such a decomposition, the equation of motion of the unknown radial part turns out as \footnote{The present WH geometry has the vanishing $T_{r\theta}$ component of the energy-momentum tensor. Furthermore, its metric form resembles that of a slowly rotating Kerr BH, and it may possess a Killing tensor~\cite{Gray:2021toe} associated with Carter's constant~\cite{Carter:1968rr}. This means that it can be considered a geometry with a separability structure~\cite{Benenti:1979erw, Demianski:1980mgt}. Consequently, separation of variables is possible for the scalar field equation in this geometry~\cite{Frolov:2006pe}.}
\be
\bea
&\frac{1}{r^2N(r)}\sqrt{1-\frac{r_0}{r}}
\frac{d}{dr}\!\left(r^{2}N(r)\sqrt{1-\frac{r_{0}}{r}}\,
\frac{dR_{l m}}{dr}
\right)+\left[
\frac{\bigl(\omega - m\,\Omega(r)\bigr)^{2}}{N(r)^{2}}
-
\frac{l(l+1)}{r^{2}}
\right]
R_{l m}(r)
=0. 
\eea
\ee
Moving forward, we find it suitable to utilize the Tortoise coordinate, defined as $dr_*/dr=1/\left(N(r)\sqrt{1-r_0/r}\right)$,
\begin{figure}[t!]
    \centering
\includegraphics[scale=0.55]{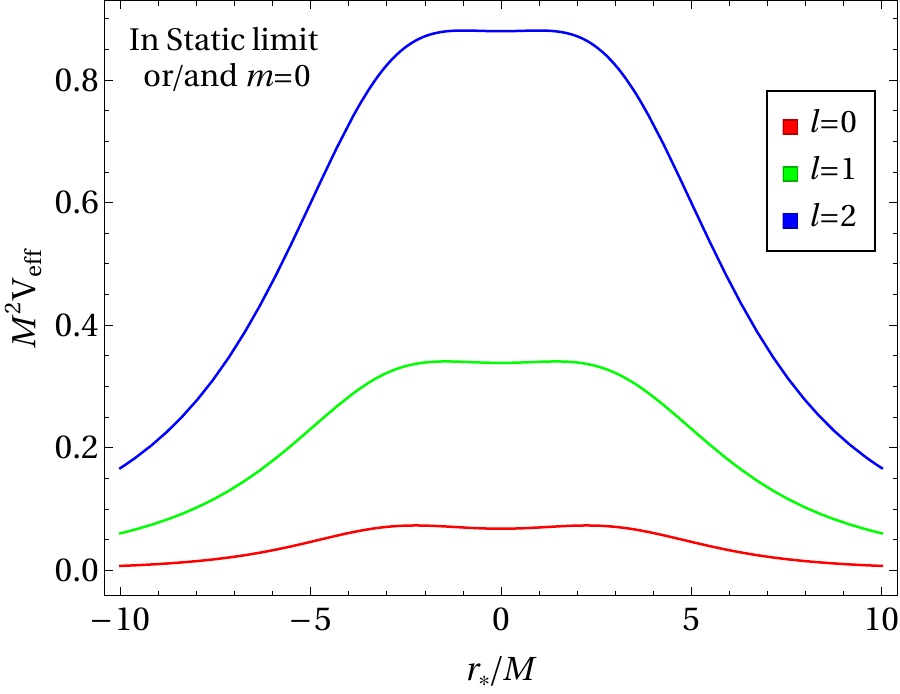}
\caption{Effective potential of the scalar field in the static limit of the WH, $a=0$, or/and for azimuthal mode, $m=0$.}\label{fig: Veff_static}
\end{figure}
\begin{figure*}[t!]
\centering
\includegraphics[width=\textwidth]{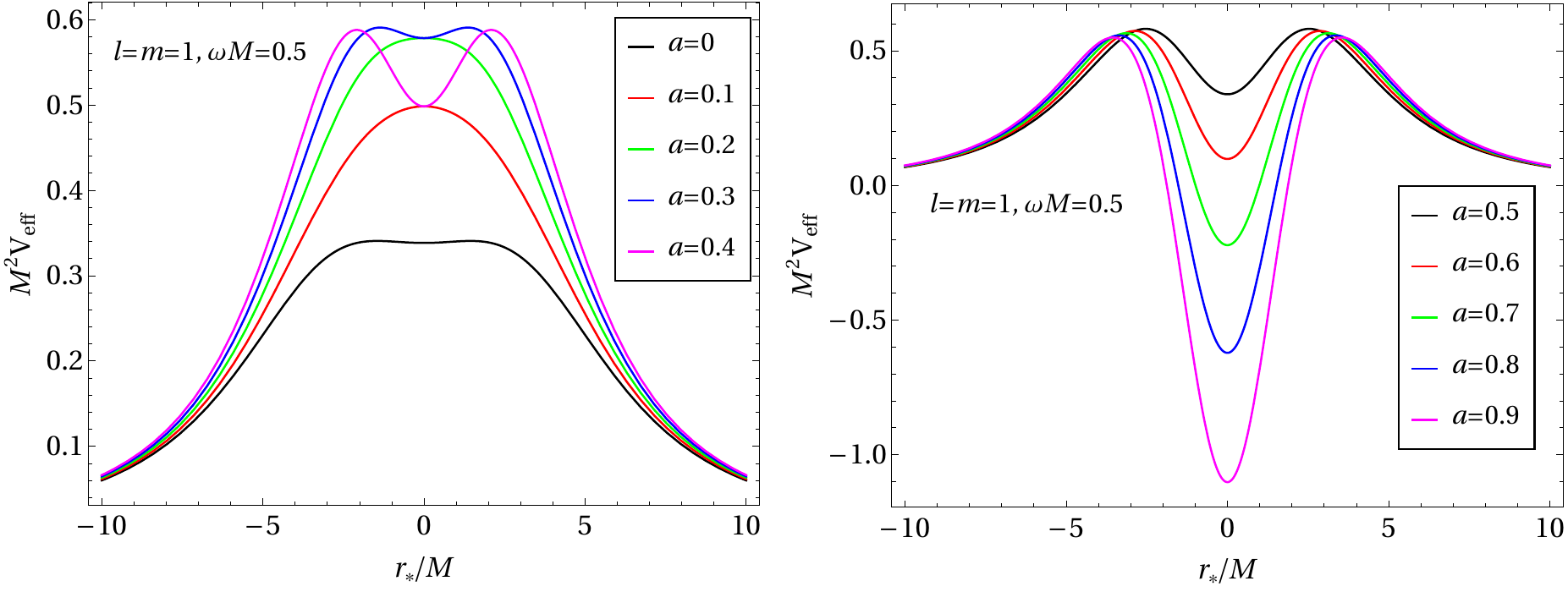}
\caption{The effective potential barrier for the scalar field has been plotted with $r_*/M$. In the {\bf left panel}, variations for different values of the spin parameter in the lower range are shown, while in the {\bf right panel}, variations for different values of the spin parameter in the higher range are presented. For plots in both panels, the frequency, orbital and azimuthal mode numbers have been kept fixed.}\label{fig: veff_diff_spin}
\end{figure*}
which smoothly connects both sides of the WH through the throat. Particularly, for $r_0\leq r< \infty$, the Tortoise radial coordinate ranges over $-\infty<r_*<\infty$, with $r_*\to 0$ as $r\to r_0$. We also find it convenient to rescale the radial part of the scalar field in the following manner
\be
R(r)\equiv\frac{u(r)}{r}.
\ee
%%%%%%%%%%%%%%%%%%%%%%%%%%%%%%%%%%%%%%%%%%%%%%%%%%%%%%%%%%%%%%%%%%%%%%%%%%%%%%%%%%%%%%
\begin{figure*}[t!]
\centering
\includegraphics[width=\textwidth]{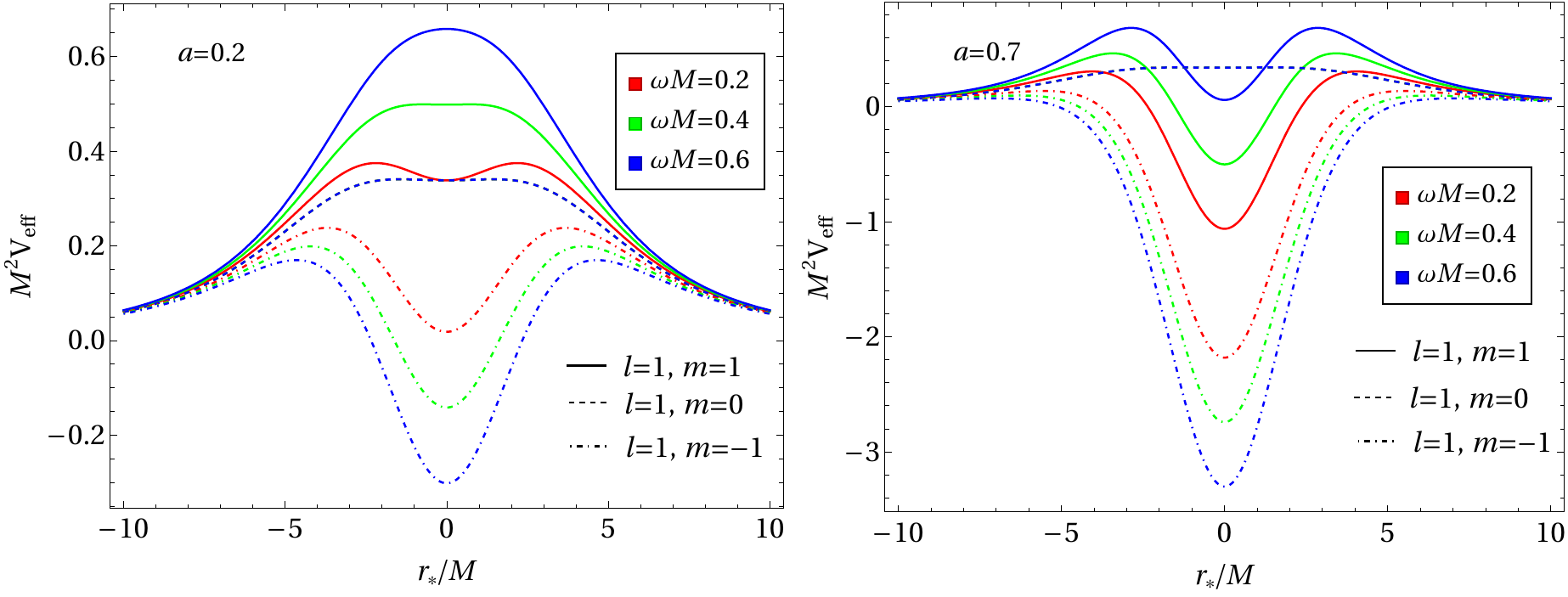}
\caption{The effective potential barrier for the scalar field has been plotted with $r_*/M$. In the {\bf left panel}, variations corresponding to different values of the scalar field frequency are shown (in different colors), along with the consideration of different azimuthal mode, $m=1$ (solid line), $m=0$ (dashed line) and $m=-1$ (dot-dashed line) for a fixed orbital modes, $l=1$, given a fixed value of the spin parameter of the WH, $a=0.3$. In the {\bf right panel}, similar variations have been considered, however, for a higher value of the spin parameter, $a=0.7$, of the WH.}\label{fig: veff_diff_omega_diff_lm}
\end{figure*}
Then, the radial equation reduces to a Schr$\ddot{o}$dinger-like form
\be\label{eq: scalar_master_eom}
\frac{d^2u}{dr^2_*}+\left[\omega^2-V_{\rm eff}(r)\right]u=0,
\ee
where the effective potential is given by
\be\label{eq: effective_pot}
\bea
&V_{\mathrm{eff}}(r)
=2m\omega\Omega(r)-m^2\Omega(r)^2+N^2(r)
\Bigg[
\frac{l(l+1)}{r^{2}}+\frac{\sqrt{1-\frac{r_0}{r}}}{rN(r)}\frac{d}{dr}\left(N(r)\sqrt{1-\frac{r_0}{r}}\right)
\Bigg].\\
\eea
\ee
This definition of the effective potential captures most of the parameter dependence of the scattering states of the scalar field quite well. In the static case $a=0$, as illustrated in Fig.~\ref{fig: Veff_static}, the height of the effective potential barrier increases monotonically with orbital mode $l$. Looking more closely, one can identify a shallow potential well bounded by barriers on both sides of the WH throat. However, as the value of $l$ increases, the well disappears. Previous studies have demonstrated that the presence of a well in the effective potential leads to trapped modes, which may manifest as resonance peaks in the transmission spectrum of static WHs, such as the generalized Ellis–Bronnikov \cite{Kar:1995jz} and the Simpson–Visser spacetime \cite{Lima:2020auu}. 

As the spin of the WH increases, shown in the left panel of Fig.~\ref{fig: veff_diff_spin}, potential wells begin to emerge more clearly. For sufficiently high spin, they develop into a pronounced potential well–type structure as illustrated in the right panel of Fig.~\ref{fig: veff_diff_spin}. Particularly, the double barrier emerges due to the symmetry of the WH spacetime on both sides of the WH throat.  On top of it, having the rotation in spacetime, the frequency and the azimuthal mode together play a significant role, as has been shown in Fig.~\ref{fig: veff_diff_omega_diff_lm}. Importantly, we find that the co-rotating modes, characterized by positive azimuthal mode $m$, exhibit only a shallow well as compared to the counter-rotating ones, with negative $m$, for smaller values of the spin, although the well becomes significantly deeper as the spin increases. Whereas the scalar modes with $m=0$ do not feel the effects of rotation, and the effective potential also becomes independent of the frequency. In the following discussion, we will see how the scattering states of the scalar field behave due to the effective potential bearing the above characteristics. 
%%%%%%%%%%%%%%%%%%%%%%%%%%%%%%%%%%%%%%%%%%%%%%%%%%%%%%%%%%%%%%%%%%%%%%%%%%%%%%%%%%%%%%
\section{Greybody factor for the scalar field}\label{sec: gbody}
In the asymptotic limits ($r_*\to \pm \infty$), the effective potential tends to vanish with respect to the frequency of the scalar field (see \eqref{eq: scalar_master_eom} and \eqref{eq: effective_pot}) due to the asymptotic flatness of the present class of WH spacetime. Hence, asymptotic radial solutions become superpositions of plane waves.
\begin{comment}
\be
\bea
&{
V_{\rm eff}(r)\simeq \left\{ % left deliminter
        \begin{array}{l} % Array with a single, Left-justified column
           ~~~~~~0,\hspace{1cm}  r\to\infty~(r_*\to \infty),\\
           
          \frac{2m\omega\Omega(r_0)}{N(r_0)^2}-\frac{m^2\Omega(r_0)^2}{N(r_0)^2},\hspace{1cm} r\to{r_0},\\
            
            ~~~~~~0, ~~~~~~~~~r\to -\infty (r_*\to -\infty)
        \end{array}
        \right.}
\eea
\ee
\end{comment}
%%%%%%%%%%%%%%%%%%%%%%%%%%%%%%%%%%%%%%%%%%%%%%%%
\begin{figure}[t!]
    \centering
\includegraphics[scale=0.55]{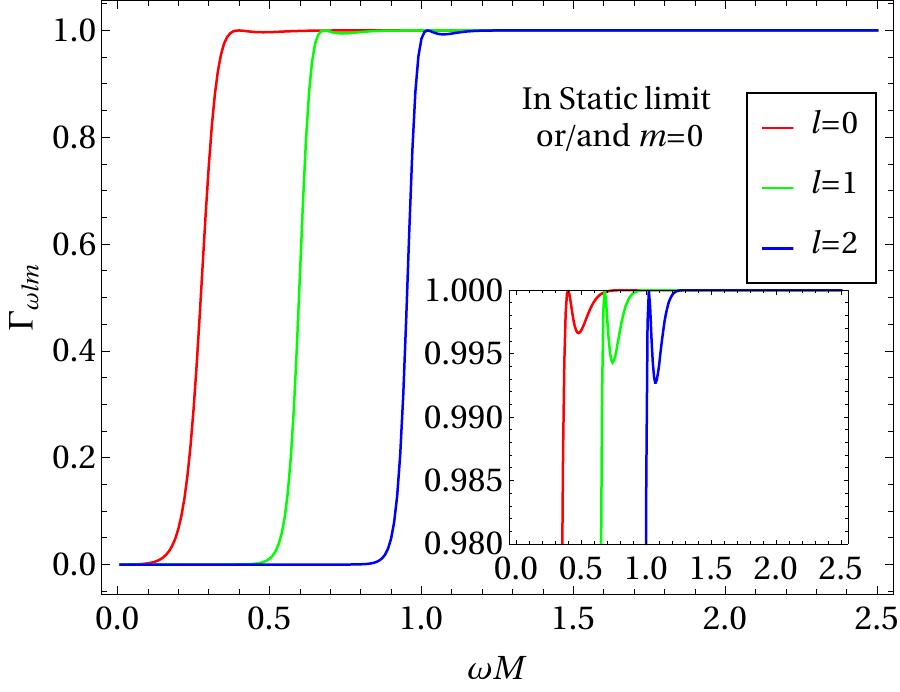}
\caption{Greybody factor for the scalar field in the static limit of the WH, $a=0$, or/and for azimuthal mode, $m=0$. The inset figure demonstrates the enlarged version before the greybody factor saturates to unity, where the appearance of resonance peaks can be seen.}\label{fig: gbody_static}
\end{figure}
%%%%%%%%%%%%%%%%%%%%%%%%%%%%%%%%%%%%%%%%%%%%%%%%
\begin{figure*}[t!]
\centering
\includegraphics[width=\textwidth]{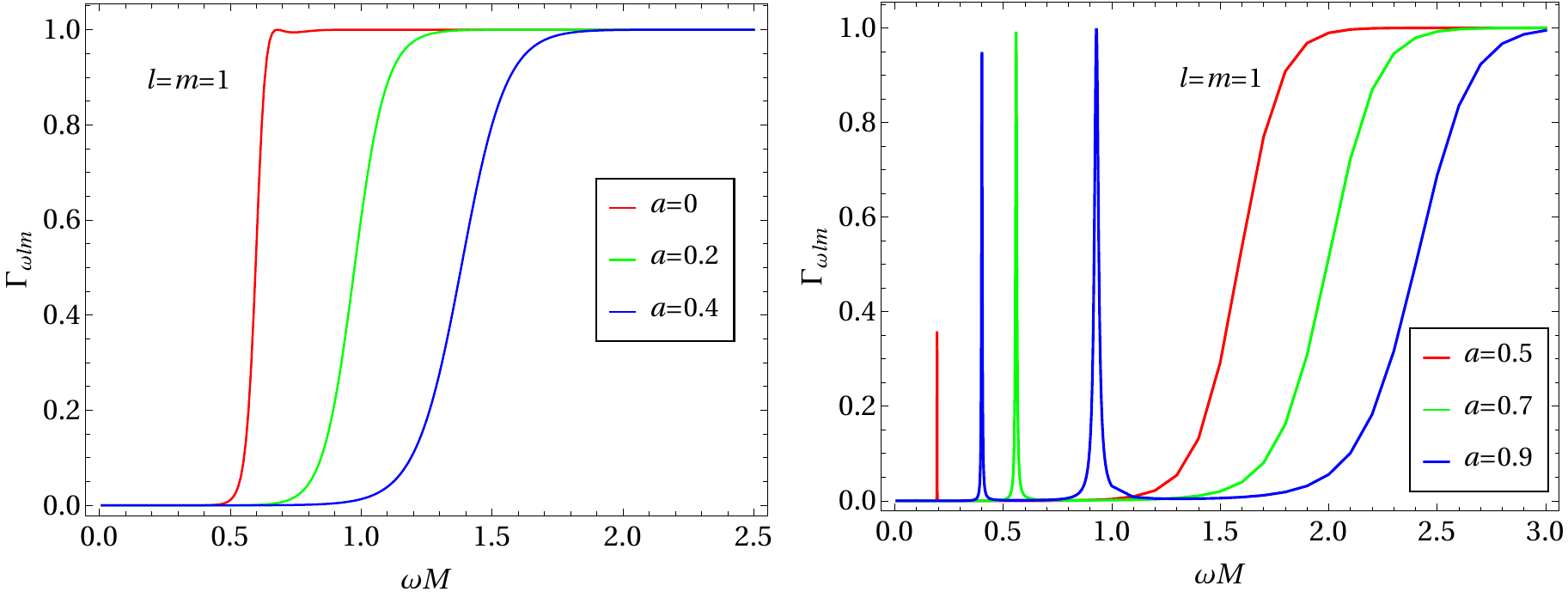}
\caption{Greybody factor for the scalar field mode, $l=1, m=1$, has been plotted with the frequency $\omega M$. In the {\bf right panel}, variations have been shown for different values of the WH spin in the lower range. In the {\bf right panel}, the behaviour is shown for higher spin.}\label{fig: gbody_diff_spin}
\end{figure*}
%%%%%%%%%%%%%%%%%%%%%%%%%%%%%%%%%%%
\begin{figure*}[t!]
\centering
\includegraphics[width=\textwidth]{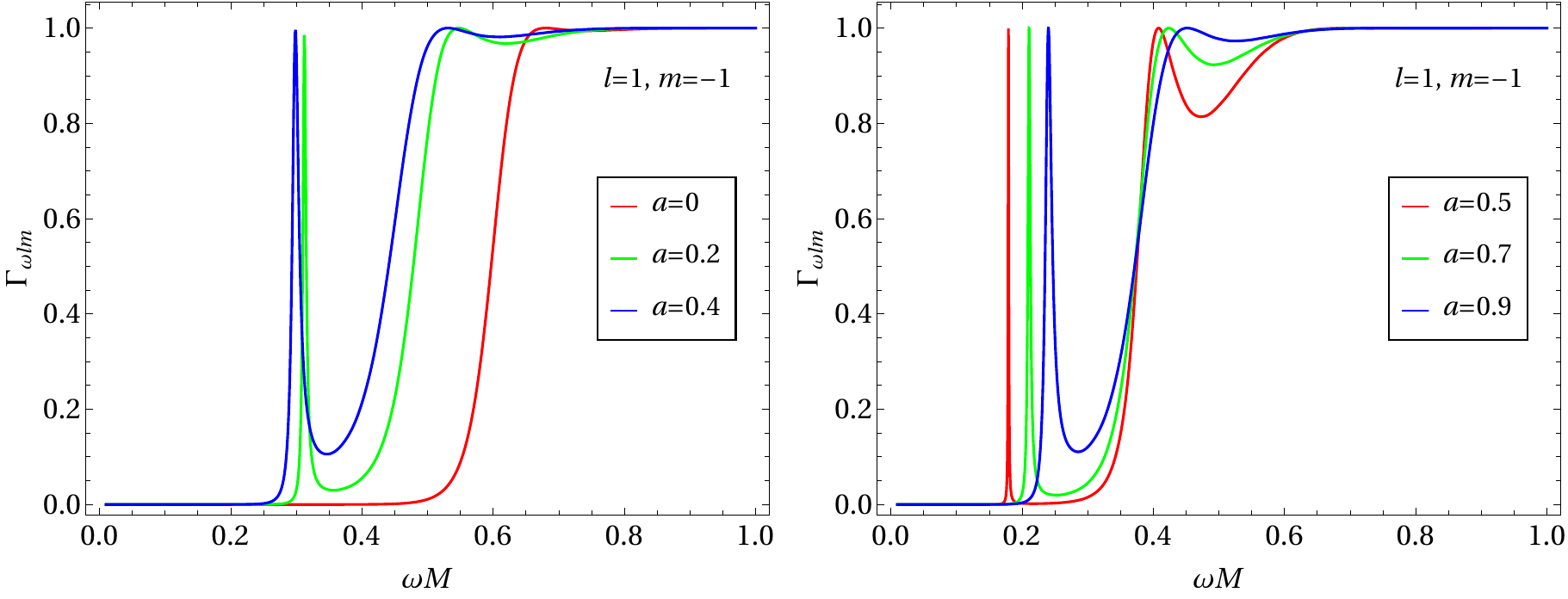}
\caption{Greybody factor for the scalar field mode, $l=1, m=-1$, has been plotted with the frequency $\omega M$. In the {\bf right panel}, variations have been shown for different values of the WH spin in the lower range. In the {\bf right panel}, the behaviour is shown for higher spin.}\label{fig: gbody_diff_spinmm1}
\end{figure*}
%%%%%%%%%%%%%%%%%%%%%%%%%%%%%%%%%%%
\begin{figure*}[t!]
\centering
\includegraphics[width=\textwidth]{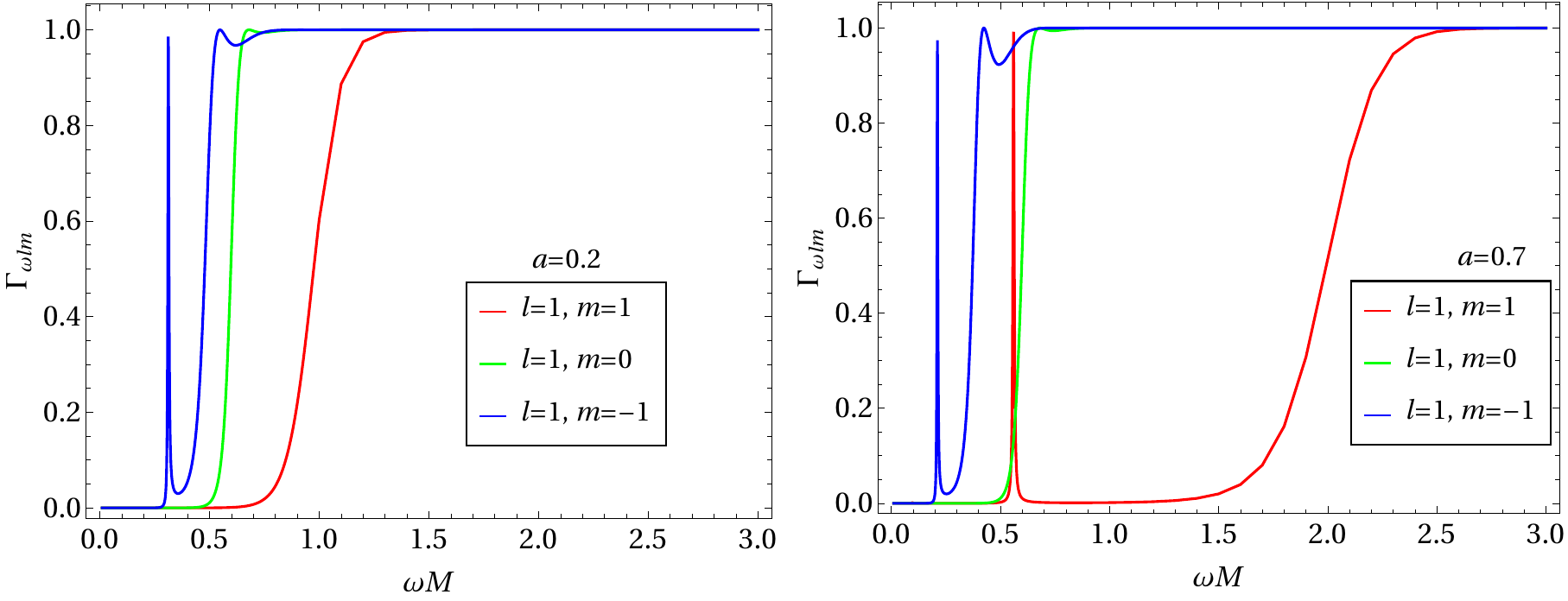}
\caption{Greybody factor for the scalar field has been plotted with the frequency $\omega M$. Variations have been shown for different azimuthal modes, $m$, corresponding to a fixed orbital mode $l=1$.}\label{fig: gbody_diff_m}
\end{figure*}
To study the scattering process by the BH spacetime, one needs to impose the boundary conditions such that the scalar wave is purely ingoing near the horizon and a superposition of ingoing and outgoing near special infinity \cite{Benone:2019all}. In contrast, the presence of a traversable throat in the WH spacetime leads the wave to be purely ingoing near $r_*\to -\infty$ \cite{Konoplya:2010kv, Lima:2020auu}. With this in mind, the asymptotic solutions can be expressed in the following manner:
\be\label{eq: u.bc}
{u(r)= \left\{ % left deliminter
        \begin{array}{l} % Array with a single, Left-justified column
           \mathcal{I}_{\omega lm} e^{-i \omega r_*}+\mathcal{R}_{\omega lm}e^{i\omega {r_*}},~~~ r_*\to{\infty},\\
          ~~~~~~~~\mathcal{T}_{\omega lm}e^{-i \omega r_*},~~~~~~~~~~~~r_*\to -\infty,
        \end{array}
        \right.}
\ee
where $\mathcal{I}_{\omega lm}, \mathcal{R}_{\omega lm}$ and $\mathcal{T}_{\omega lm}$ represent the incident, reflection, and transmission coefficients, respectively. Whereas, $e^{i\omega {r_*}}$ denotes the outgoing mode, while $e^{-i\omega {r_*}}$ denotes the ingoing mode, with the temporal part given in \eqref{eq: mode_decomposition}. The conserved Wronskian associated with the linear second-order differential equation \eqref{eq: scalar_master_eom}, $W[u,u^*]=u\pr_{r_*}u^{*}-u^*\pr_{r_*}u$, turns out to be
\be
W[u,u^*]= {\left\{ % left deliminter
        \begin{array}{l} % Array with a single, Left-justified column
           2i\omega(|\mathcal{I}_{\omega lm}|^2-|\mathcal{R}_{\omega lm} |^2),~~r_*\to \infty,\\
          ~~~~~2i\omega|\mathcal{T}_{\omega lm}|^2,~~~~~~~~~~~~~r_*\to -\infty.\\
        \end{array}
        \right.}
\ee
Then the conservation of the Wronskian leads to the following condition based on the above equation, 
\be\label{eq: flux_conserv}
(|\mathcal{I}_{\omega lm}|^2-|\mathcal{R}_{\omega lm}|^2)=|\mathcal{T}_{\omega lm} |^2.
\ee
This condition implies the absence of superradiance phenomena in the WH spacetime, even in the presence of rotation, which is in contrast to the usual rotating BH spacetime \cite{Benone:2019all, Karmakar:2024uzt}. In a traversable WH, any negative-energy component generated inside the ergoregion is not irreversibly lost behind a horizon, unlike in a BH spacetime \cite{Brito:2015oca}. Instead, the wave can propagate through the throat and emerge into the other asymptotic region \cite{Konoplya:2010kv}. The incident wave will be partly reflected and partly transmitted at the throat without any amplification. Nevertheless, in this scenario, one can suitably define the greybody factor, accounting for the transmission of the matter field \cite{Rosato:2025byu}, to study the scattering phenomena, as
\be\label{eq: gbody_defn}
\Gamma_{\omega lm}=1-\Bigg|\frac{\mathcal{R}_{\omega lm}}{\mathcal{I}_{\omega lm}}\Bigg|^2 = \frac{|\mathcal{T}_{\omega lm} |^2}{|\mathcal{I}_{\omega lm} |^2}.
\ee
%This factor is also called transmittivity and is defined as the absolute value squared of the transmission and incident coefficients~.
Therefore, the greybody factor represents the probability that an incident wave will successfully cross the potential barrier and through the throat of a WH. In what follows, we will outline the procedure for computing the reflection and transmission coefficients numerically, which will determine the greybody factor, as defined above.
\subsection{Numerical recipe for the computation of the greybody factor} 
Although in a rotating BH spacetime, the reflection and transmission coefficients can be analytically determined in the low- and high-frequency limit \cite{Benone:2019all, Macedo:2013afa}, we find it simpler to deduce these coefficients numerically in the presence of the double barrier potential and perform the analysis for arbitrary frequencies. For this purpose, the master equation \eqref{eq: scalar_master_eom} that governs the scalar field is solved in Mathematica using the built-in NDSolve function, explicitly setting the initial condition to near negative spatial infinity $r_{*-\infty}=-100 M$. To solve the linear second-order differential equation, two conditions are needed, for which we choose the derivative of the field at $r_{*-\infty}=-100 M$ to provide another initial condition.  It is important to note that this condition is not an independent condition and carries the same ingoing requirement. Therefore, the set of boundary conditions all reads $[u(r), u'(r)]_{r_*\to -\infty}$, which can be straightforwardly deduced from \eqref{eq: u.bc}. Then the numerical solution is obtained at $r_\infty=100 M$, i.e., far from the WH throat. The stability of the final result with respect to the variation of this outer boundary will be discussed below. Nevertheless,  once the solution is obtained, it is matched to the asymptotic analytical form \eqref{eq: u.bc}, and consequently, the reflection and incident coefficients are also determined numerically. Afterwards, substituting these coefficients into the above definition \eqref{eq: gbody_defn}, the greybody factor was numerically evaluated for a range of frequencies with resolution $\Delta(\omega M)\sim 5\times 10^{-5}$, fixing the accuracy and precision up to $\sim 15$ decimal places.

The Teo's WH in the absence of rotation, as mentioned before, boils down to the form of a Morris-Thorne-type traversable WH. In this limiting case, the appearance of a shallow well in the effective potential for the massless scalar field has been discussed in the previous section. 
In Fig.~\ref{fig: gbody_static}, we have illustrated the behavior of the greybody factor for this case. The distinctive characteristic of resonant transmission can be seen for different orbital modes $l$, before the corresponding greybody factors saturate to the maximum value $\Gamma=1$. The resonance occurs for a certain frequency for each orbital mode, which can be understood following the effective potential, for example, with $\omega M \simeq 0.5$, for $l=0$, the value of $\omega^2M^2$ cuts through the double peak of the potential well (see Fig.~\ref{fig: Veff_static}) and causes the traveling wave outside the potential barrier to be temporarily trapped as it falls into the shallow well. Such temporarily confined oscillations are associated with QNMs~\cite{Macedo:2018yoi}, whose properties will be analyzed in detail in \cref{subsec: long_lived_QNMs}. Owing to the finite height of the potential barriers, the confined waves can eventually escape through quantum-mechanical tunneling~\cite{sakurai1986modern}. During their confinement, repeated reflections between the barriers give rise to multiple interfering wave components. When the appropriate phase-matching condition is satisfied, constructive interference enhances the transmission probability, producing sharp spectral peaks characteristic of Breit–Wigner-type resonances~\cite{Lima:2020auu, Macedo:2018yoi}.
\begin{comment}
Moreover, the transmission spectrum around the resonant transmission can be well approximated by the Breit-Wigner formula,
\be
\Gamma(\omega)=\frac{A_{\omega lm}}{(\omega-\omega_R)^2+\omega^2_I},
\ee
where $A_{\omega lm}$ represents the constant factor dependent on the mode. Whereas $\omega_R$ is the real part of the resonant frequency corresponding to the peak position in the greybody factor, and $\omega_I$ stands for the spectral width. As mentioned above, these frequencies are also related to the real and imaginary parts of the long-lived complex QNMs \cite{Lima:2020auu, Macedo:2018yoi}.
\end{comment}

Once rotation is introduced in the WH spacetime, the resonant transmission becomes more pronounced and depends on the azimuthal mode $m$. Therefore, we first present the behavior of the greybody factor for co-rotating modes, $l=1, m=1$, in Fig.~\ref{fig: gbody_diff_spin}. Interestingly, resonant transmission, although present in a slight manner in the static case, does not occur for this mode until $a \simeq 0.6$. This closely follows from the behavior of the frequency-dependent effective potential for the WH spin in the lower range (see Fig.~\ref{fig: veff_diff_spin}), where a slight increase in the spin away from $a=0$ leads to the disappearance of the potential well structure. The well then re-emerges at $a=0.3$. However, for resonant transmission, the scalar mode should be bounded by the barrier, $\omega^2< V_{\rm eff}$, and should remain as traveling waves within the well, $\omega^2>V_{\rm eff}$. Such a scenario is viable for higher spin for the co-rotating modes (see the right panel of Fig.~\ref{fig: veff_diff_spin}), therefore, resonant peaks are appearing in the right panel of Fig.~\ref{fig: gbody_diff_spin}. For the lower-spin, one might expect that resonant transmission could be achieved by increasing the frequency; however, as shown in Fig.~\ref{fig: veff_diff_omega_diff_lm}, for $l=1, m=1$ the barrier height increases with frequency, thus suppressing the possibility of resonant transmission. 

For the counter rotating modes, as shown in Fig.~\ref{fig: gbody_diff_spinmm1}, for $l=1, m=-1$ we see that resonant transmission always occurs because the rotation lowers the corresponding effective potential much more efficiently \eqref{eq: effective_pot}. For a fixed orbital mode, $l=1$, the comparison of the greybody factor for different azimuthal modes, $m$, is presented in Fig.~\ref{fig: gbody_diff_m}. Since an ergoregion appears in the equatorial plane for \(a\gtrsim 0.184\), it is instructive to investigate the stability of these resonant modes (frequencies) with both positive and negative azimuthal numbers in light of the ergoregion instability mentioned in the Introduction. Importantly, in all the cases considered above, the frequencies corresponding to the resonant peaks are expected to be closely associated with the real parts of long-lived QNM frequencies~\cite{Lima:2020auu, Macedo:2018yoi}, as will be investigated in detail in \cref{subsec: long_lived_QNMs}. The emergence of these long-lived modes is a generic feature of horizonless compact objects \cite{Franzin:2022iai} and carries important implications for the ringdown waveform \cite{Rosato:2025byu, Rosato:2025lxb, Biswas:2023ofz}.

Before proceeding to the QNM analysis, we first establish the numerical convergence of the transmission spectrum with respect to the computational domain, frequency resolution, numerical precision, accuracy goals, and other relevant computational parameters.
%%%%%%%%%%%%%%%%%%%%%%%%%%%%%%%%%%%%%%%%%%%%%

{\it Varying the boundary location} ($r_{*\pm\infty}$): To assess the convergence of the numerical solutions, we examine the sensitivity of the transmission spectrum to variations in the asymptotic boundary locations. For the representative counter-rotating mode \((l,m)=(1,-1)\), we track the convergence of the resonance peak for \(a=0.2\) and \(a=0.7\). As shown in Table~\ref{tab: diff_bounday}, the peak transmission remains stable under successive outward shifts of the boundaries, with variations below \(10^{-5}\). The corresponding resonance frequencies are also insensitive to these changes at the adopted frequency resolution, \(\Delta(\omega M)=5\times10^{-5}\).
\begin{table}[h]
\centering
\begin{tabular}{|c|c|c||c|c|c|}
\hline
\multicolumn{3}{|c||}{$a=0.2$} &
\multicolumn{3}{c|}{$a=0.7$} \\
\hline
$|r_{*\pm\infty}/M|$ & 
$\Gamma^{\omega lm}_{\rm peak}$ & &
$|r_{*\pm\infty}/M|$ &
$\Gamma^{\omega lm}_{\rm peak}$ & \\
\hline
\cline{1-2}\cline{4-6}
100  & 0.99999737& & 100  & 0.99999868& \\
\cline{1-2}\cline{4-6}
150   & 0.99999587& & 150 & 0.99999789& \\
\cline{1-2}\cline{4-6}
200 & 0.9999954& & 200 & 0.99999999& \\
\cline{1-2}\cline{4-6}
250& 0.99999522 & &250 & 0.99999995& \\
\cline{1-2}\cline{4-6}
300 & 0.99999515 & & 300 & 0.99999968& \\
\cline{1-2}\cline{4-6}
350 & 0.99999513 & & 350 & 0.99999998 &\\ 
\hline
\end{tabular}
\caption{Convergence of the resonant transmission spectrum by varying the asymptotic boundary, $r_{*\pm \infty}$, for a fixed mode $l=1,\, m=-1$. The representative table is given for two different values of the spin parameter.}\label{tab: diff_bounday}
\end{table}

%%%%%%%%%%%%%%%%%%%%%%%%%%%%%%%%%%
{\it Frequency resolution}: We have investigated the convergence of the resonant transmission spectrum with respect to the frequency resolution, $\Delta(\omega M)$. As shown in Table \ref{tab: freq_resol}, lower-resolution frequency grids underestimate the maximum transmission. Upon progressively refining the frequency resolution, both the resonance frequency and the peak transmission converge rapidly, with the results becoming unchanged for \(\Delta(\omega M)\simeq 5\times10^{-5}\).
\begin{table}[h]
\centering
\begin{tabular}{|c|c|c|c||c|c|c|c|}
\hline
\multicolumn{4}{|c||}{$a=0.2$} &
\multicolumn{4}{c|}{$a=0.7$} \\
\hline
$\Delta(\omega M)$ & $M\omega_{\rm peak}$ & 
$\Gamma^{\omega lm}_{\rm peak}$ & &
$\Delta(\omega M)$ & $M\omega_{\rm peak}$ &
$\Gamma^{\omega lm}_{\rm peak}$ & \\
\hline
0.01   & 0.31 & 0.57716785 & & 0.01   & 0.21 & 0.91382193& \\
\cline{1-3}\cline{5-7}
0.005  & 0.31 & 0.57716785 & & 0.005  &0.21 & 0.91382193& \\
\cline{1-3}\cline{5-7}
0.001  & 0.312 & 0.9856838 & & 0.001  & 0.211& 0.97411502& \\
\cline{1-3}\cline{5-7}
0.0005 & 0.3125 & 0.99715793 & & 0.0005 & 0.2105 & 0.99524853& \\
\cline{1-3}\cline{5-7}
0.0001 & 0.3123 & 0.99975324 & & 0.0001 & 0.2106 &0.99950709& \\
\cline{1-3}\cline{5-7}
0.00005 & 0.31235 & 0.99999737& & 0.00005 & 0.21065 & 0.99999868& \\
\cline{1-3}\cline{5-7}
0.00001 & 0.31235 & 0.99999737& & 0.00001 & 0.21065 & 0.99999868& \\
\hline
\end{tabular}
\caption{Convergence of the resonant transmission spectrum under successive frequency refinements for a fixed mode $l=1,\, m=-1$. The representative table is given for two different values of the spin parameter.}\label{tab: freq_resol}
\end{table}
%%%%%%%%%%%%%%%%%%%%%%%%%%%%%%%

\begin{figure}[t!]
    \centering
\includegraphics[scale=0.5]{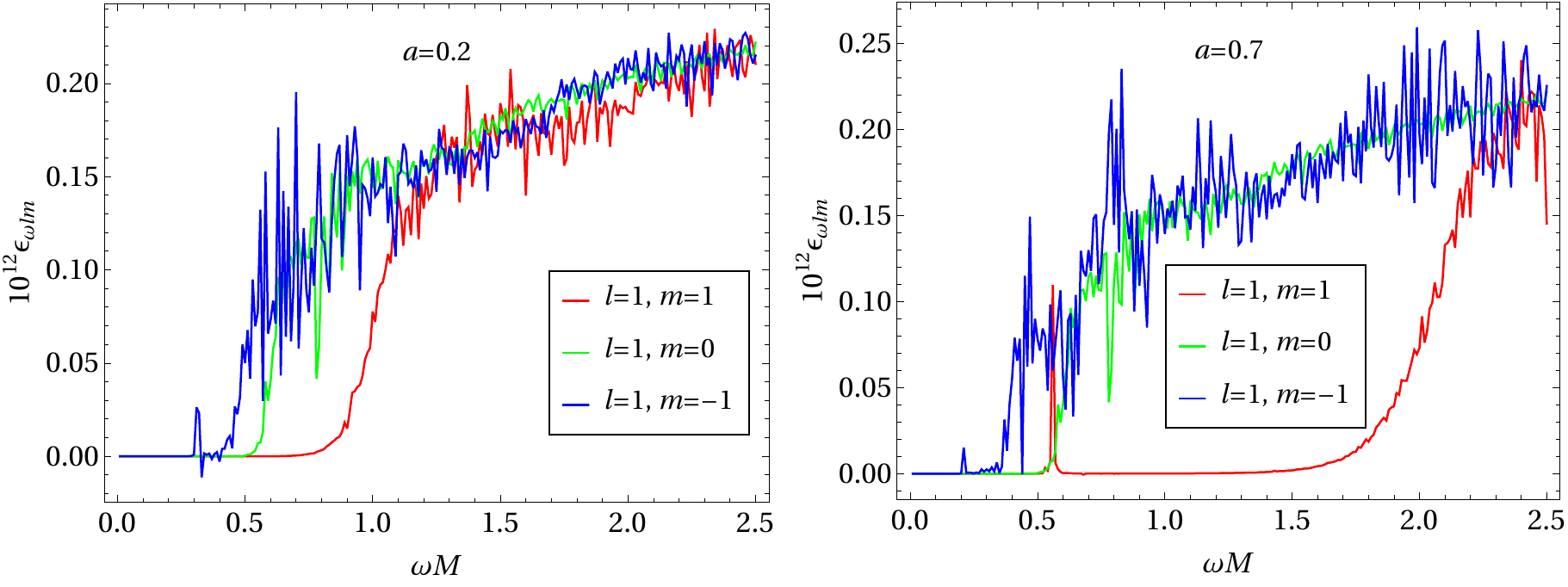}
\caption{The flux-conservation deviation parameter is plotted as a function of the scalar-field frequency for \(a=0.2\) (left panel) and \(a=0.7\) (right panel), with fixed \(l=1\) and \(m=1,0,-1\).}\label{fig: epsilon_l1diffm}
\end{figure}

{\it Numerical validation of the flux conservation:} To further assess the numerical accuracy of our calculations, we verify the conservation of the scalar field flux, and thereby the conservation of the Wronskian, which leads to the formula given in \eqref{eq: flux_conserv}. We therefore define the deviation parameter as
\be
\epsilon_{\omega lm}=\frac{|\mathcal{T}_{\omega lm} |^2}{|\mathcal{I}_{\omega lm} |^2}+\Bigg|\frac{\mathcal{R}_{\omega lm}}{\mathcal{I}_{\omega lm}}\Bigg|^2-1,
\ee
which should vanish for an accurate numerical solution. The above deviation parameter has been numerically evaluated, and the result has been shown in Fig.~\ref{fig: epsilon_l1diffm}. The numerical deviations depend on the precision and accuracy adopted in the computation. For a numerical precision and accuracy up to 15 decimal digits, we find $\epsilon_{\omega lm}\sim 10^{-14}$. This demonstrates that the deviations remain negligibly small over the relevant frequency range, confirming that the numerical solutions satisfy the flux-conservation relation with high accuracy for all the modes considered.
We should also mention that the Mathematica code developed for this purpose successfully reproduces the existing scattering results in the case of static WH \cite{LimaJunior:2022zvu}.
%%%%%%%%%%%%%%%%%%%%%%%%%%%%%%%%%%%%%%%%%%%%%%%%
%%%%%%%%%%%%%%%%%%%%%%%%%%%%%%%%%%%%%%%%%%%%%%%%
\subsection{Long-lived quasinormal modes}\label{subsec: long_lived_QNMs}
The ingoing modes can be trapped for a certain amount of time and therefore referred to as trapped modes, which in the complex domain are related to the long-lived QNMs.  The boundary condition for the QNMs in the asymptotically flat WH spacetime is given as follows \cite{Konoplya:2018ala, Konoplya:2025hgp}:
\be\label{eq: qnm_wh_bc}
{u(r)= \left\{ % left deliminter
        \begin{array}{l} % Array with a single, Left-justified column
          e^{i\omega {r_*}},~~~ r_*\to{\infty},\\
          e^{-i \omega r_*},~~~r_*\to -\infty.
        \end{array}
        \right.}
\ee
In contrast to the BH case, the QNM boundary condition here requires purely outgoing waves at both spatial infinities. Since \(r_*\to\pm\infty\) corresponds to the two asymptotic regions, the homogeneous radial equation and the outgoing-wave boundary conditions take the same form at both ends, allowing the radial solutions to be classified according to their reflection symmetry about the throat \cite{Cardoso:2016rao}. Consequently, the problem can be reduced to the semi-infinite domain \(r_*\in[0,\infty)\) by imposing the appropriate throat boundary condition \cite{Konoplya:2025hgp}: \(\left. du/dr_*\right|_{r_*=0}=0\) for the even (symmetric) sector and \(u(r_*=0)=0\) for the odd (antisymmetric) sector. In this setup, we compute the complex QNM frequencies, \(\omega=\omega_R+i\omega_I\), for different spin parameters and orbital modes using the direct integration method~\cite{Lima:2020auu, Chandrasekhar:1975zza}. The frequencies are determined numerically using Mathematica's FindRoot function, with initial guesses for the complex frequencies guided by the corresponding resonant peaks in the transmission spectrum. The resulting long lived QNM frequencies are presented in Table~\ref{tab:qnm_scalar}. 
\begin{table}[h]
\centering
\begin{tabular}{|l|c|c|c|c|c|c|}
\hline
Spin & Orbital mode & \multicolumn{2}{c|}{Even sector} &
\multicolumn{2}{c|}{Odd sector} &
$M\Delta\omega_{\rm FWHM}$ \\
\cline{3-6}
$a$ & $(l,m)$ & $M\omega_R$ & $-M\omega_I$ &
$M\omega_R$ & $-M\omega_I$ & ($\simeq-2M\omega_I$) \\
\hline

\multirow{3}{*}{0} & $(0,\forall m)$ & 0.34374 & 0.038750 &
--- & --- & 0.077500\\
\cline{2-7}

& $(1,\forall m)$ & 0.60528 & 0.052383 &
--- & --- & 0.104766\\
\cline{2-7}

& $(2,\forall m)$ & 0.95997 & 0.04736 &
--- & --- & 0.094720\\
\hline

\multirow{2}{*}{0.2} & \multirow{2}{*}{$(1,-1)$} & 0.31677 & 0.0017068 &
--- & --- & 0.0034136\\
\cline{3-7}

& & --- & --- & 0.50419 & 0.051701 &
0.103402\\
\hline
\multirow{2}{*}{0.4} & \multirow{2}{*}{$(1,-1)$} & --- & --- & 0.30780 & 0.0038570 & 0.0077140\\
\cline{3-7}

& & 0.47650 & 0.063876 & --- & --- & 0.127752
\\
\hline
\multirow{3}{*}{0.5} & $(1,1)$ & 0.19115 & $-0.000043894$ &
--- & --- & ---\\
\cline{2-7}
& \multirow{2}{*}{$(1,-1)$} & --- & --- &
0.18797 & 0.00027713 & 0.00055426\\
\cline{3-7}

& & 0.40291 & 0.027981 & --- & --- & 0.055962\\
\hline

\multirow{3}{*}{0.7} & $(1,1)$ & 0.55534 & $-0.0028452$ &
--- & --- & ---\\
\cline{2-7}

& \multirow{2}{*}{$(1,-1)$} & 0.22290 & 0.0012147 &
--- & --- & 0.0024294\\
\cline{3-7}

& & --- & --- & 0.40582 & 0.041496 &
0.082992\\
\hline

\multirow{4}{*}{0.9} & \multirow{2}{*}{$(1,1)$} & --- & --- &
0.39231 & $-0.0012763$ & ---\\
\cline{3-7}

& & 0.92513 & $-0.010782$ &
--- & --- & ---\\
\cline{2-7}
& \multirow{2}{*}{$(1,-1)$} & --- & --- &
0.25508 & 0.0037853 & 0.0075706\\
\cline{3-7}
& & 0.41286 & 0.058990 & --- & --- & 0.11798 \\
\hline
\end{tabular}

\caption{QNM frequencies of scalar perturbations in the rotating Teo WH, classified into even and odd
parity sectors. The full width at half maximum is estimated
from the imaginary part of the QNM frequency as $\Delta\omega_{\rm FWHM}=-2M\omega_I$ for damped modes.
Entries marked by dashes in the FWHM column correspond
to unstable modes with $\omega_I>0$, while in the rest of the columns they stand for the absence of the corresponding long-lived stable or unstable modes.}

\label{tab:qnm_scalar}
\end{table}
For the adopted time dependence \(e^{-i\omega t}\), modes with \(\omega_I<0\) are damped, whereas those with \(\omega_I>0\) exhibit exponential growth, signaling an instability of the scalar perturbation. For an isolated, long-lived QNM, the real part \(\omega_R\) approximately determines the location of the corresponding Breit–Wigner resonance in the transmission spectrum, while the imaginary part governs its linewidth and decay timescale \cite{Lima:2020auu}. In particular, the full width at half maximum is estimated as \(\Delta\omega_{\rm FWHM}\simeq-2\omega_I\), with the corresponding lifetime given by \(\tau\simeq1/|\omega_I|\) for damped modes \cite{Maggio:2017ivp}. The dimensionless linewidths, \(M\Delta\omega_{\rm FWHM}\), computed from the QNM frequencies are also listed in the table. Notably, the appearance of modes with \(\omega_I>0\), particularly in the corotating sector, indicates the presence of an ergoregion instability, as also discussed in earlier studies of rotating WHs \cite{Franzin:2022iai},  for which the usual interpretation in terms of a positive resonance linewidth is not applicable. In contrast, the stable counter rotating modes exhibit long lived oscillations with small negative imaginary parts, giving rise to narrow resonant features in the transmission spectrum. The dimensionless linewidths, \(M\Delta\omega_{\rm FWHM}\), computed from the QNM frequencies are also listed in the table. The corresponding dimensionless damping and growth times, defined respectively as \(\tau_{\rm damp}/M=1/(-M\omega_I)\) for \(\omega_I<0\) and \(\tau_{\rm grow}/M=1/(M\omega_I)\) for \(\omega_I>0\), provide further insight into the stability of the resonant modes. For the stable counter-rotating mode with \(a=0.7\) and \((l,m)=(1,-1)\), we obtain \(\tau_{\rm damp}/M\simeq823.25\), indicating a long-lived resonance. In contrast, the corotating mode with \(a=0.7\) and \((l,m)=(1,1)\) has \(\tau_{\rm grow}/M\simeq351.47\), while the corresponding mode at \(a=0.9\) exhibits a growth time of \(\tau_{\rm grow}/M\simeq783.52\) in the odd sector. These results illustrate that the instability timescale depends on the spin and parity of the mode, and establish the connection between the QNM spectrum and resonant transmission for the rotating traversable WH spacetime.
%%%%%%%%%%%%%%%%%%%%%%%%%%%%%%%%%%%%%%%%%%%%%%%%
\section{Normalization of the scalar field}\label{sec:  norm_planewave}
To keep the present analysis on general footing, the incoming massless scalar wave is considered to have an incident angle with the WH spin axis $\gamma$. Now, the Rayleigh expansion provides for the decomposition of a plane wave in terms of partial waves, which can be expressed as \cite{Unruh:1976fm},
\be\label{eq: rayleigh_expansion}
e^{-i\omega\left(t+r \cos\theta'\right)}=\sum_l(2l+1) (-i)^le^{-i\omega t}j_l(\omega r)P_l(\cos\theta'),
\ee
with 
\be
\cos\theta'=\cos\gamma \cos\theta+\sin\gamma\sin\theta \cos\phi,
\ee
where $j_l(\omega r)$ represents the spherical Bessel function and $P_l(\cos\theta')$ denotes the Legendre function \cite{abramowitz1964handbook}.
Whereas $\theta'$ stands for the projected angular coordinate of WH with the incident angle, so that
\be
P_l(\cos\theta')=\frac{4\pi}{2l+1}\sum^l_{m=-l}Y_{lm}(\theta, \varphi)Y^*_{lm}(\gamma, 0).
\ee
Inserting this formula into \eqref{eq: rayleigh_expansion}, the plane wave decomposition can be rewritten as
\be
e^{-i\omega\left(t+r \cos\theta'\right)}=\sum_{lm} 4\pi (-i)^le^{-i\omega t}j_l(\omega r)Y_{lm}(\theta, \varphi)Y^*_{lm}(\gamma, 0),
\ee
To this end we utilize the fact that for $r\to\infty$ or $\omega r>>1$, the asymptotic form of the spherical Bessel function reads 
\be
j_l(\omega r)\sim i^{l+1}\frac{e^{-i\omega r}}{2\omega r}+(-i)^{l+1}\frac{e^{i\omega r}}{2\omega r}.
\ee
Substituting this form into the preceding plane wave expansion, we arrive at
\be\label{eq: in_plane_wave}
e^{-i\omega\left(t+r \cos\theta'\right)}=\frac{2\pi}{\omega r}\sum_{lm} (-i)^le^{-i\omega t}\left[i^{l+1}e^{-i\omega r}+(-i)^{l+1}e^{i\omega r}\right]Y_{lm}(\theta, \varphi)Y^*_{lm}(\gamma, 0),
\ee
Simplifying the expansion further, we get
\be\label{eq: in_plane_wave}
e^{-i\omega\left(t+r \cos\theta'\right)}=\sum_{lm}2\pi i\left[\frac{e^{-i\omega\left( t+r\right)}}{\omega r}-(-1)^l\frac{e^{-i\omega\left( t-r\right)}}{\omega r}\right]Y_{lm}(\theta, \varphi)Y^*_{lm}(\gamma, 0),
\ee
On the other hand, using the asymptotic form of the radial solution at spatial infinity \eqref{eq: u.bc}, the asymptotic form of the scalar field \eqref{eq: mode_decomposition} can be expressed as
\be\label{eq: asymp_phi}
\phi=\sum_{lm}\mathcal{N}_{\omega lm}\left[\mathcal{I}_{\omega lm}\frac{e^{-i\omega\left(t+r\right)}}{r}+\mathcal{R}_{\omega lm}\frac{e^{-i\omega\left(t-r\right)}}{r}\right] Y_{lm}(\theta, \varphi).
\ee
Since the ingoing component represents the incident radiation, whereas the outgoing component contains the scattered contributions, we determine the overall normalization by matching only the ingoing spherical-wave amplitudes \eqref{eq: in_plane_wave} to have a unit amplitude incident plane wave. Hence, the overall normalization constant is fixed to be
\be\label{eq: norm_factor}
\mathcal{N}_{\omega lm}=i\frac{2\pi}{\omega\mathcal{I}_{\omega lm}}Y^*_{lm}(\gamma, 0).
\ee
In the next section, this normalization factor will be utilized to define and determine the partial transmission cross section of the WH for the scalar field.
\begin{figure}[t!]
    \centering
\includegraphics[scale=0.55]{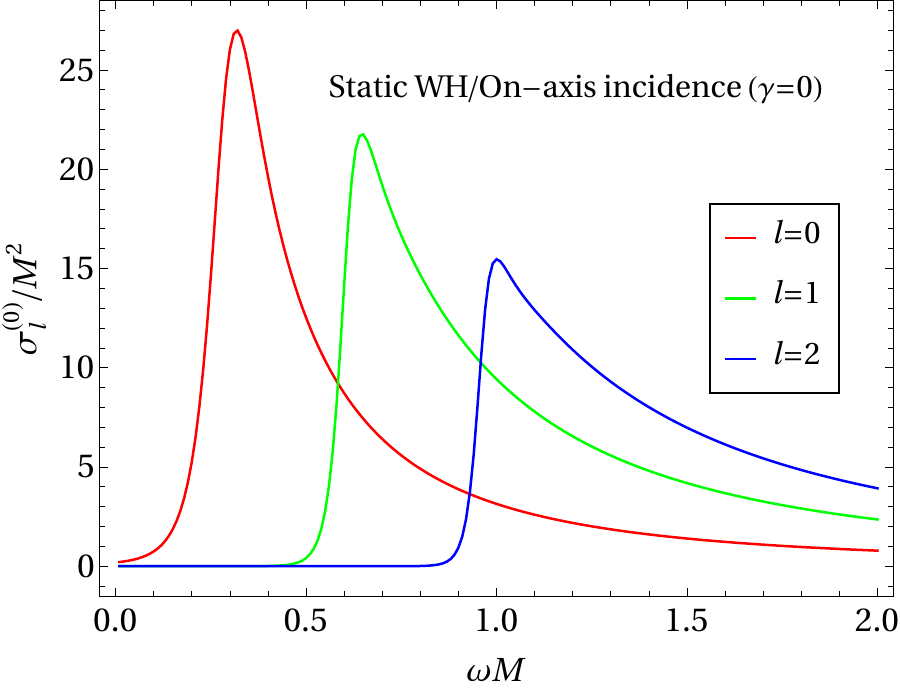}
\caption{The partial transmission cross section for a scalar field is plotted in the static WH limit. Notably, the same result holds for on-axis incidence, $\gamma=0$ and $m=0$, for which rotation decouples from the radial dynamics.}\label{fig: pcs_static}
\end{figure}
%%%%%%%%%%%%%%%%%%%%%%%%%%%%%%
\begin{figure*}[t!]
\centering
\includegraphics[width=\textwidth]{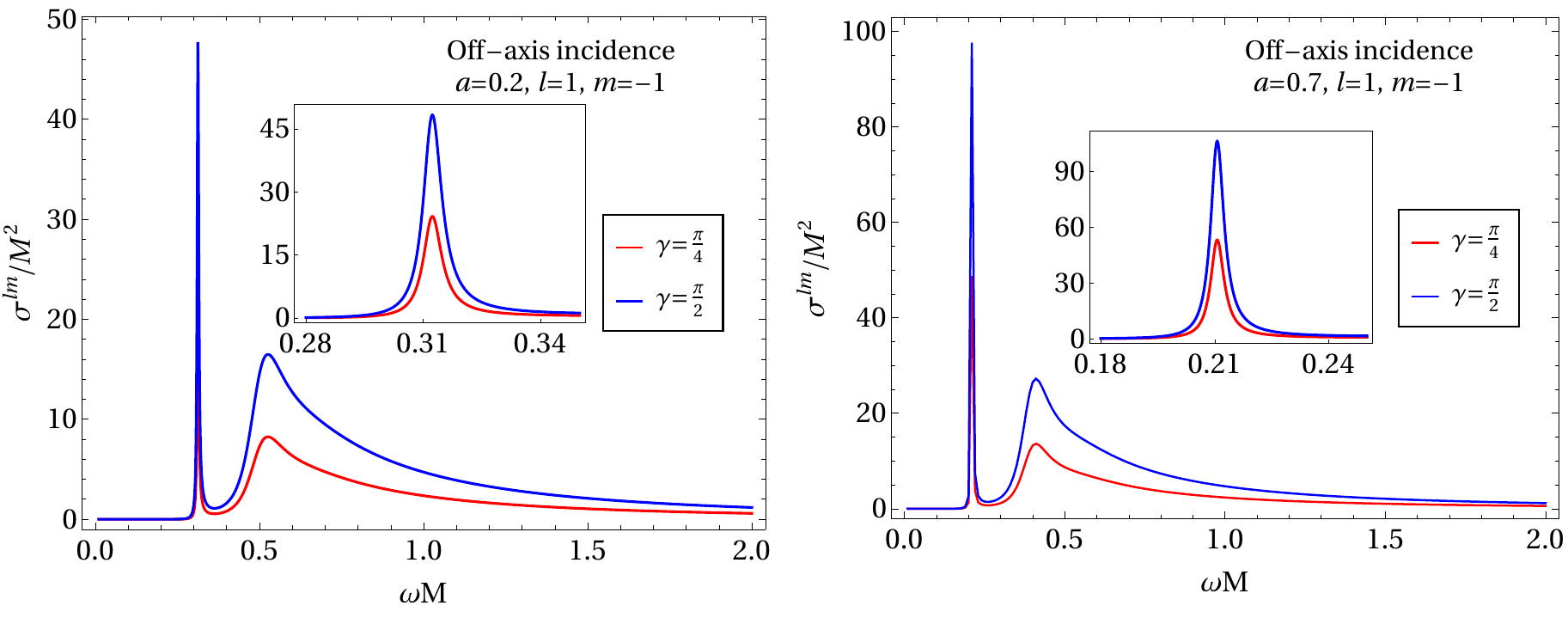}
\caption{The partial transmission cross section of the WH has been plotted with respect to the frequency of the scalar field. {\bf In the left panel:} We have considered smaller values for the spin parameter, $a=0.2$, whereas, {\bf in the right panel} a higher value, $a=0.7$, has been chosen. In each panel, two branches are shown for two different angles of incidence for the fixed stable scalar-field mode $l=1$, $m=-1$. The inset in each panel shows an enlarged view of the spectrum near the resonance.}\label{fig: pcs_rotating}
\end{figure*}
\begin{figure*}[t!]
\centering
\includegraphics[width=\textwidth]{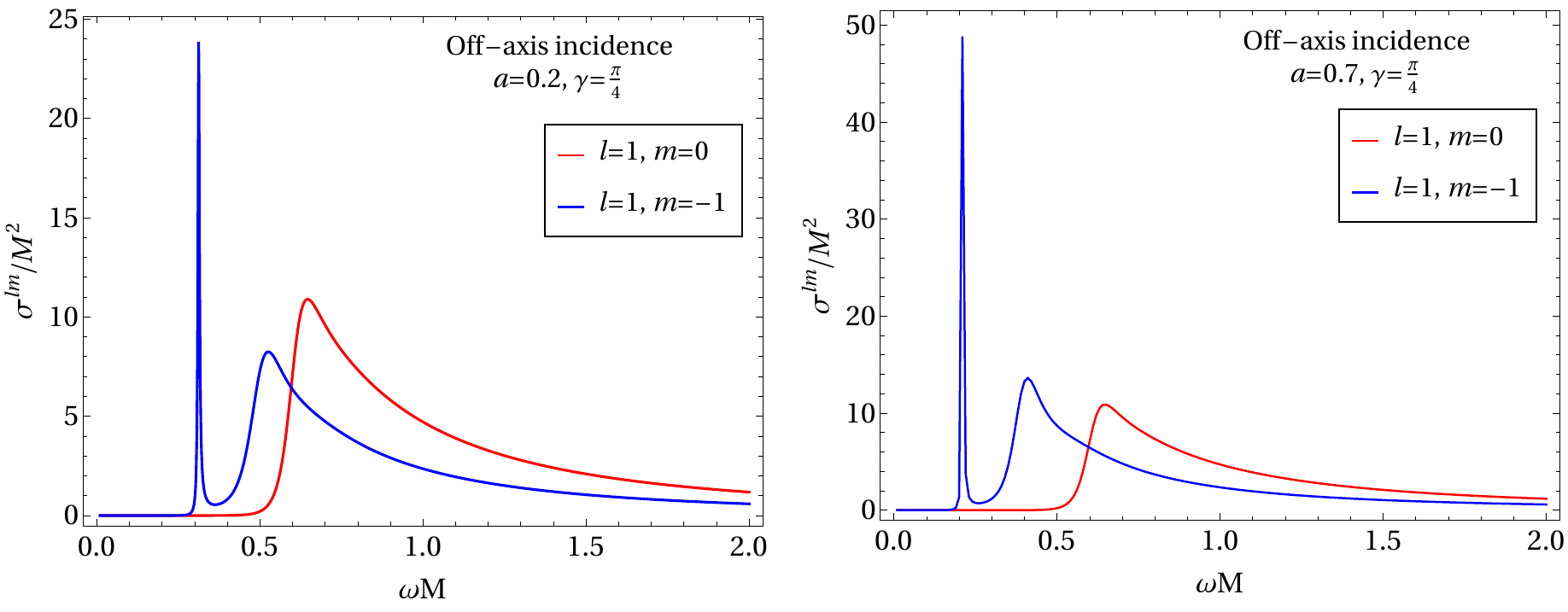}
\caption{The partial transmission cross section of the WH has been plotted with respect to the frequency of the scalar field. {\bf In the left panel:} We have considered smaller values for the spin parameter, $a=0.2$, whereas, {\bf in the right panel} a higher value, $a=0.7$, has been chosen. In each panel, two branches corresponding to the stable scalar-field modes $l=1, m=-1$ and $l=1, m=0$ are compared at a fixed angle of incidence, $\gamma=\pi/4$.}\label{fig: pcs_rotating_diffm}
\end{figure*}
\section{Partial Transmission Cross Sections for Stable Modes}\label{sec: defn_pcs}
As discussed in the previous section, the presence of ergoregion instabilities can affect the physical interpretation of stationary scattering processes. We therefore restrict our analysis to the partial transmission cross sections of selected scalar modes that remain stable within the parameter range investigated, allowing us to explore their mode-dependent resonant features. Since a traversable WH possesses neither an event horizon nor a dissipative boundary, the incident scalar-field flux is not absorbed but instead propagates through the throat into the second asymptotic region. Accordingly, the partial transmission cross section characterizes the fraction of the incident scalar-field flux transmitted through the WH throat into the second asymptotic region for each angular mode.

The energy flux measured by a static observer (as Teo's WH is asymptotically static) at $r\to\infty$, per unit time is
\be
\pr_t\mathcal{E}=\int{d}\Omega{r^2}\mathcal{T}_{tr},
\ee
where for a massless scalar field the stress energy tensor $\mathcal{T}_{\mu\nu}$ is given by
\be
\mathcal{T}_{\mu\nu}=\frac{1}{2}(\pr_\mu\phi^*\pr_\nu\phi+\pr_\mu\phi\pr_\nu\phi^*)-\frac{1}{2}g_{\mu\nu}\pr^\alpha\phi^*\pr_\alpha\phi.
\ee
Substituting the asymptotic form of the massless scalar field solution \eqref{eq: asymp_phi} 
\be
\pr_t\mathcal{E}=\int{d}\Omega{r^2}\mathcal{T}_{tr}=\sum_{l}|\mathcal{N}_{\omega lm}|^2\omega^2\left(|\mathcal{I}|^2-|\mathcal{R}|^2\right).
\ee
The transmission cross section, analogous to the absorption cross
section for BHs, can be defined as the ratio of the transmitted
energy flux rate to the incident energy flux density
\cite{Cardoso:2019dte}, which, with the adopted normalization,
is given by $\omega^2$. Accordingly, the transmission cross
section takes the form
\be
\sigma=\frac{\partial_t\mathcal{E}}{\omega^2}.
\ee
Substituting the normalization factor from
\eqref{eq: norm_factor}, the cross section can be expressed
as a sum over the individual partial-wave contributions,
\be\label{eq: defn_pcs}
\sigma=\sum_{lm}\sigma_{lm},
\qquad
\sigma_{lm}
=\frac{4\pi^2}{\omega^2}
|Y_{lm}(\gamma,0)|^2\Gamma_{\omega lm},
\ee
where $\sigma_{lm}$ denotes the partial transmission
cross section associated with the angular mode $(l,m)$. For on-axis incidence, i.e., $\gamma=0$, the partial transmission cross section simplifies to
\be\label{eq: on_axis_norm}
\sigma_{lm}^{(0)}
=\frac{\pi}{\omega^2}(2l+1)
\delta_{m0}\Gamma_{\omega lm}.
\ee
Having established the numerical procedure for computing
the greybody factor in \Cref{sec: gbody}, we can now
determine the corresponding partial transmission cross
sections for the scalar field. As evident from the above
expressions, the partial cross section
differs from the greybody factor for a given mode only
through a frequency-dependent multiplicative factor.
Consequently, the resonant features identified in the
transmission spectrum are also reflected in the
corresponding partial cross sections. Although a total transmission cross section can be defined for static WHs by summing over all the contributing angular modes, its interpretation in the rotating case requires additional care because of the ergoregion instability associated with certain
co-rotating modes, as demonstrated in \cref{subsec: long_lived_QNMs}. Therefore, to maintain a consistent comparison between static and rotating configurations while restricting the analysis to dynamically stable modes, we focus exclusively on the
partial transmission cross sections in what follows.

In Fig.~\ref{fig: pcs_static}, we show the partial transmission cross section for a scalar field in the static WH limit. The weak resonant feature identified in the corresponding greybody factor (see Fig.~\ref{fig: gbody_static}) manifests as a slightly sharper rise in the cross section, compared with the smoother increases and decreases observed in the BH case~\cite{Benone:2014qaa}. A very similar result was obtained long ago for a generalized Ellis-Bronnikov WH spacetime \cite{Kar:1995jz}. Moreover, for incidence along the rotation axis ($\gamma=0$), only $m=0$ modes contribute to the transmission cross section. Since the spin dependence in \eqref{eq: scalar_master_eom} enters solely through $m\Omega$, the radial equation for these modes is independent of rotation. Consequently, the on-axis transmission cross section coincides with that of the static WH, unlike the Kerr BH case, where the $m=0$ radial equation retains an explicit spin dependence~\cite{Benone:2019all}.  

Because of the rotation, the spectrum is expected to depend strongly on the angle at which the incident wave approaches the WH throat relative to its spin axis. In Fig.~\ref{fig: pcs_rotating} and Fig.~\ref{fig: pcs_rotating_diffm}, we have shown the behavior of the partial transmission cross section for the rotating scenario for off-axis incidence. We find that the amplitude of the resonance, particularly in the lower range of the frequency, increases and becomes sharper with increasing rotation. Although, the magnitude in the corresponding greybody factor (see Fig.~\ref{fig: gbody_diff_m}) for the two panels in Fig.~\ref{fig: pcs_rotating} seems to be similar, the resonance frequency being lower for higher rotation the enhancement in the magnitude happens in the cross section \eqref{eq: defn_pcs}. The spectrum also exhibits stronger resonance with the increase in the incident angle from the rotation axis, and the maximum effect is found for the perpendicular incidence. Concerning Fig.~\ref{fig: pcs_rotating_diffm} it is important to note that, for the particular Teo-class WH considered here, the greybody factor associated with the $m=0$ modes is independent of the rotation parameter and coincides with its static counterpart. However, the corresponding partial transmission cross section retains a dependence on the incidence angle $\gamma$ through the angular factor $|Y_{l0}(\gamma,0)|^2$. This dependence arises from the projection of the incident plane wave onto the individual angular modes rather than from the rotation of the WH. In particular, for perpendicular incidence, $\gamma=\pi/2$, the partial transmission cross section vanishes for odd $l$ with $m=0$, even though the corresponding greybody factor remains nonzero. This signifies that the incident scalar wave has no projection onto these particular angular modes, so they do not contribute to the transmitted flux, even though the WH itself remains capable of transmitting scalar waves in those modes. 
%%%%%%%%%%%%%%%%%%%%%%%%%%%%%%%%%%
\section{Conclusion and outlook}\label{sec: concl}
In the present work, we have investigated the propagation of a massless
scalar field in a rotating traversable WH spacetime, focusing on the
greybody factor, resonant transmission, and QNM spectrum. We find that
rotation significantly influences the transmission spectrum, producing
sharp resonant peaks at characteristic frequencies. The spectrum also
depends strongly on the angle of incidence relative to the rotation
axis. However, due to the particular structure of the Teo-class WH
considered here, on-axis incidence involves only $m=0$ modes, for
which the rotational contribution vanishes. Consequently, the
transmission spectrum coincides with that of the static configuration,
in contrast to the Kerr BH case~\cite{Macedo:2013afa}. As the incidence
angle increases, the resonant features become more prominent, reaching
their maximum strength for perpendicular incidence.

In the static limit, we find no pronounced resonant peaks in the
transmission spectrum. This differs from the behavior of
Simpson--Visser WHs~\cite{LimaJunior:2022zvu} and clean
photon-sphere objects (ClePhOs)~\cite{Macedo:2018yoi}, where sharp
resonances can arise even without rotation due to wave trapping
within the effective-potential cavity. For the present WH, the
static effective potential has a relatively shallow cavity near
the throat~\cite{Kar:1995jz, Riley:2026avn, Konoplya:2010kv}.
Rotation modifies this structure, producing a deeper cavity,
particularly for counter-rotating modes, and thereby giving rise
to sharp resonances at low frequencies.

Unlike rotating BHs, where superradiance allows
the extraction of rotational energy under suitable conditions
\cite{Brito:2015oca}, the rotating WH considered here does not
exhibit superradiant amplification in scalar wave scattering.
Instead, rotation gives rise to resonant transmission, providing
a distinct scattering behavior compared with rotating BHs \cite{Benone:2019all}. The QNM analysis further reveals a close connection between
the resonant transmission peaks and the real parts of long lived
QNM frequencies~\cite{Lima:2020auu, Macedo:2018yoi}. In particular,
the narrow resonances associated with stable counter rotating
modes correspond to weakly damped QNMs with small imaginary
parts. On the other hand, co-rotating modes develop
positive imaginary parts, indicating an ergoregion instability, consistent with previous findings for rotating Kerr-like WHs~\cite{Franzin:2022iai}.

We have also computed the partial transmission cross sections
for the stable modes. Unlike the absorption cross section of
a BH, the transmission cross section of a WH measures the
scalar wave flux passing through the throat into the second
asymptotic region, rather than being absorbed by the object.
Although a total transmission cross section can be defined
for static WHs, the presence of unstable modes in rotating
configurations complicates its interpretation. We therefore
restrict our comparison to partial transmission cross sections
associated with stable modes, which retain the characteristic
resonant features of the greybody factors.

These results may be relevant for distinguishing rotating
WHs from BHs. In particular, recent studies have shown that
greybody factors can leave characteristic imprints on
gravitational wave ringdown signals
\cite{Rosato:2025byu, Rosato:2025lxb, Biswas:2023ofz}.
Although the present analysis concerns a scalar field,
which is often used as a proxy for higher-spin perturbations, similar resonant features may also arise for electromagnetic and gravitational waves. However, establishing their observational significance requires a direct analysis of these perturbations and their associated waveforms.

Finally, we have considered a particular Teo-class WH,
chosen for its simple structure. Other configurations
within this class~\cite{Gyulchev:2018fmd}, as well as
alternative rotating traversable WH geometries
\cite{Kleihaus:2014dla, Chew:2016epf, Cisterna:2023uqf},
may exhibit different transmission and stability properties. In particular, extending the present investigation of resonant transmission to rotating Kerr-like WHs~\cite{Mazza:2021rgq}, whose QNM spectra and stability properties have already been studied \cite{Franzin:2022iai}, would provide an interesting opportunity to explore the connection between scattering resonances and QNMs in these geometries. Further extending the present analysis to these geometries, particularly for electromagnetic and gravitational perturbations, would help identify the features that distinguish rotating WHs from their BH counterparts.

%However, insofar as both are asymptotically flat, spatially large limit behavior of the scattering states, however, is quite similar; the only difference it makes is in the two-sided  case of WHs.Comparing several concepts and characteristics appeared in the geometry of black holes or wormholes
%The greybody factor represents the probability that Hawking radiation generated near a black hole's event horizon being transmitted to infinity over the potential barrier. Consequently, when Hawking radiation generated near the horizon is observed in the asymptotic region, it is modified~\cite{Page:1976df}. 

%In a Kerr black hole, the presence of a horizon and negative energy states in the ergosphere allow a sign change in $\omega_{H}$ within the transmission solution part of equation (11). In contrast, Teo's rotating wormhole lacks a horizon and there is no possibility of a sign change in $\omega_{\infty}$ within the transmission solution part, and so on... 

%It would be helpful to explain the differences and characteristics observed in the geometries of static black holes, Kerr black holes, static wormholes, and Teo's rotating wormholes with regard to the boundary conditions, Greybody factor and absorption cross section, citing relevant reference papers.

%%%%%%%%%%%%%%%%%%%%%%%%%%%%%%%%%%
%%%%%%%%%%%%%%%%%%%%%%%%%%%%%%%%%%

\section*{Acknowledgements} %%%%%%%%
RK thanks the research group of Xian-Hui Ge for the useful discussions held during the weekly meetings at the Physics Department of Shanghai University. RK is also grateful to Debaprasad Maity for numerous discussions
over the years on scattering phenomena in curved backgrounds in general. 
BHL was supported by the National Research Foundation of Korea (NRF)  (RS-2026-25473640 and RS-2024-00339204) and by Overseas Visiting Fellow Program of Shanghai University. BHL would like to thank the hospitality of the KIAS and APCTP during the visit where a part of this work has been done . WL was supported by RS-2026-25484780, BHL and WL were also supported by Basic Science Research Program of the National Research Foundation of Korea funded by the Ministry of Education through CQUeST (RS-2020-NR049598). 
%%%%%%%%%%%%%%%%%%%%%%%%%%%%%%%%%%%%%%%%%
%%%%%%%%%%%%%%%%%%%%%%%%%%%%%%%%%%%%
\bibliographystyle{JHEP}
\bibliography{ref}

@article{Einstein:1935tc,
    author = "Einstein, Albert and Rosen, N.",
    title = "{The Particle Problem in the General Theory of Relativity}",
    doi = "10.1103/PhysRev.48.73",
    journal = "Phys. Rev.",
    volume = "48",
    pages = "73--77",
    year = "1935"
}

@article{Misner:1957mt,
    author = "Misner, Charles W. and Wheeler, John A.",
    title = "{Classical physics as geometry: Gravitation, electromagnetism, unquantized charge, and mass as properties of curved empty space}",
    doi = "10.1016/0003-4916(57)90049-0",
    journal = "Annals Phys.",
    volume = "2",
    pages = "525--603",
    year = "1957"
}

@article{Ellis:1973yv,
    author = "Ellis, H. G.",
    title = "{Ether flow through a drainhole - a particle model in general relativity}",
    doi = "10.1063/1.1666161",
    journal = "J. Math. Phys.",
    volume = "14",
    pages = "104--118",
    year = "1973"
}

@article{Chetouani:1984qdm,
    author = "Chetouani, Lyazid and Cl{\'e}ment, G{\'e}rard",
    title = "{Geometrical optics in the Ellis geometry}",
    doi = "10.1007/BF00762440",
    journal = "Gen. Rel. Grav.",
    volume = "16",
    number = "2",
    pages = "111--119",
    year = "1984"
}

@article{Morris:1988cz,
    author = "Morris, M. S. and Thorne, K. S.",
    title = "{Wormholes in space-time and their use for interstellar travel: A tool for teaching general relativity}",
    doi = "10.1119/1.15620",
    journal = "Am. J. Phys.",
    volume = "56",
    pages = "395--412",
    year = "1988"
}

@book{Visser:1995cc,
    author = "Visser, Matt",
    title = "{Lorentzian wormholes: From Einstein to Hawking}",
    isbn = "978-1-56396-653-8",
    year = "1995"
}

@article{Lobo:2005us,
    author = "Lobo, Francisco S. N.",
    title = "{Phantom energy traversable wormholes}",
    eprint = "gr-qc/0502099",
    archivePrefix = "arXiv",
    doi = "10.1103/PhysRevD.71.084011",
    journal = "Phys. Rev. D",
    volume = "71",
    pages = "084011",
    year = "2005"
}

@article{Halder:2019urh,
    author = "Halder, Shibaji and Bhattacharya, Subhra and Chakraborty, Subenoy",
    title = "{Spherically symmetric wormhole solutions in a general anisotropic matter field}",
    eprint = "1903.03343",
    archivePrefix = "arXiv",
    primaryClass = "gr-qc",
    doi = "10.1016/j.physletb.2019.02.041",
    journal = "Phys. Lett. B",
    volume = "791",
    pages = "270--275",
    year = "2019"
}

@article{Kim:2019ojs,
    author = "Kim, Hyeong-Chan and Lee, Youngone",
    title = "{Spherically Symmetric Wormholes with anisotropic matter}",
    eprint = "1905.10050",
    archivePrefix = "arXiv",
    primaryClass = "gr-qc",
    doi = "10.1088/1475-7516/2019/09/001",
    journal = "JCAP",
    volume = "09",
    pages = "001",
    year = "2019"
}

@article{DeFalco:2020afv,
    author = "De Falco, Vittorio and Battista, Emmanuele and Capozziello, Salvatore and De Laurentis, Mariafelicia",
    title = "{General relativistic Poynting-Robertson effect to diagnose wormholes existence: static and spherically symmetric case}",
    eprint = "2004.14849",
    archivePrefix = "arXiv",
    primaryClass = "gr-qc",
    doi = "10.1103/PhysRevD.101.104037",
    journal = "Phys. Rev. D",
    volume = "101",
    number = "10",
    pages = "104037",
    year = "2020"
}

@article{Hassan:2021egb,
    author = "Hassan, Zinnat and Mandal, Sanjay and Sahoo, P. K.",
    title = "{Traversable Wormhole Geometries in Gravity}",
    eprint = "2102.00915",
    archivePrefix = "arXiv",
    primaryClass = "gr-qc",
    doi = "10.1002/prop.202100023",
    journal = "Fortsch. Phys.",
    volume = "69",
    number = "6",
    pages = "2100023",
    year = "2021"
}

@article{Bouhmadi-Lopez:2021zwt,
    author = "Bouhmadi-L{\'o}pez, Mariam and Chen, Che-Yu and Chew, Xiao Yan and Ong, Yen Chin and Yeom, Dong-han",
    title = "{Traversable wormhole in Einstein 3-form theory with self-interacting potential}",
    eprint = "2108.07302",
    archivePrefix = "arXiv",
    primaryClass = "gr-qc",
    doi = "10.1088/1475-7516/2021/10/059",
    journal = "JCAP",
    volume = "10",
    pages = "059",
    year = "2021"
}

@article{Kim:2024mam,
    author = "Kim, Hyeong-Chan and Kim, Sung-Won and Lee, Bum-Hoon and Lee, Wonwoo",
    title = "{Charged traversable wormholes: charge without charge}",
    eprint = "2405.10013",
    archivePrefix = "arXiv",
    primaryClass = "gr-qc",
    reportNumber = "CQUeST-2024-0734",
    doi = "10.1007/s40042-025-01515-2",
    journal = "J. Korean Phys. Soc.",
    volume = "88",
    number = "4",
    pages = "401--417",
    year = "2026"
}

@article{Hochberg:1997wp,
    author = "Hochberg, David and Visser, Matt",
    title = "{Geometric structure of the generic static traversable wormhole throat}",
    eprint = "gr-qc/9704082",
    archivePrefix = "arXiv",
    doi = "10.1103/PhysRevD.56.4745",
    journal = "Phys. Rev. D",
    volume = "56",
    pages = "4745--4755",
    year = "1997"
}

@article{Kim:2013tsa,
    author = "Kim, Sung-Won",
    title = "{Flare-out condition of a Morris-Thorne wormhole and finiteness of pressure}",
    eprint = "1302.3337",
    archivePrefix = "arXiv",
    primaryClass = "gr-qc",
    doi = "10.3938/jkps.63.1887",
    journal = "J. Korean Phys. Soc.",
    volume = "63",
    pages = "1887--1891",
    year = "2013"
}

@article{Koga:2025bqw,
    author = "Koga, Yasutaka and Maeda, Ryota and Saito, Daiki and Uemichi, Keiya and Yoshida, Daisuke",
    title = "{Dynamical formation of charged wormholes}",
    eprint = "2505.20040",
    archivePrefix = "arXiv",
    primaryClass = "gr-qc",
    reportNumber = "KUNS-3056, YITP-25-85, NU-QG-7",
    doi = "10.1103/mtcj-13wm",
    journal = "Phys. Rev. D",
    volume = "113",
    number = "4",
    pages = "044035",
    year = "2026"
}

@article{Kim:1997jf,
    author = "Kim, Sung-Won and Kim, Sang Pyo",
    title = "{The Traversable wormhole with classical scalar fields}",
    eprint = "gr-qc/9907012",
    archivePrefix = "arXiv",
    doi = "10.1103/PhysRevD.58.087703",
    journal = "Phys. Rev. D",
    volume = "58",
    pages = "087703",
    year = "1998"
}

@article{Lobo:2009ip,
    author = "Lobo, Francisco S. N. and Oliveira, Miguel A.",
    title = "{Wormhole geometries in f(R) modified theories of gravity}",
    eprint = "0909.5539",
    archivePrefix = "arXiv",
    primaryClass = "gr-qc",
    doi = "10.1103/PhysRevD.80.104012",
    journal = "Phys. Rev. D",
    volume = "80",
    pages = "104012",
    year = "2009"
}

@article{Kanti:2011jz,
    author = "Kanti, Panagiota and Kleihaus, Burkhard and Kunz, Jutta",
    title = "{Wormholes in Dilatonic Einstein-Gauss-Bonnet Theory}",
    eprint = "1108.3003",
    archivePrefix = "arXiv",
    primaryClass = "gr-qc",
    doi = "10.1103/PhysRevLett.107.271101",
    journal = "Phys. Rev. Lett.",
    volume = "107",
    pages = "271101",
    year = "2011"
}

@article{Kim:2016pky,
    author = "Kim, Jin Young and Park, Mu-In",
    title = "{On a new approach for constructing wormholes in Einstein{\textendash}Born{\textendash}Infeld gravity}",
    eprint = "1608.00445",
    archivePrefix = "arXiv",
    primaryClass = "hep-th",
    doi = "10.1140/epjc/s10052-016-4497-7",
    journal = "Eur. Phys. J. C",
    volume = "76",
    number = "11",
    pages = "621",
    year = "2016"
}

@article{Ovgun:2018xys,
    author = {{\"O}vg{\"u}n, Ali and Jusufi, Kimet and Sakall{\i}, {\.I}zzet},
    title = "{Exact traversable wormhole solution in bumblebee gravity}",
    eprint = "1804.09911",
    archivePrefix = "arXiv",
    primaryClass = "gr-qc",
    doi = "10.1103/PhysRevD.99.024042",
    journal = "Phys. Rev. D",
    volume = "99",
    number = "2",
    pages = "024042",
    year = "2019"
}

@article{Huang:2019arj,
    author = "Huang, Hyat and Yang, Jinbo",
    title = "{Charged Ellis Wormhole and Black Bounce}",
    eprint = "1909.04603",
    archivePrefix = "arXiv",
    primaryClass = "gr-qc",
    doi = "10.1103/PhysRevD.100.124063",
    journal = "Phys. Rev. D",
    volume = "100",
    number = "12",
    pages = "124063",
    year = "2019"
}

@article{Godani:2020gbr,
    author = "Godani, Nisha and Samanta, Gauranga C.",
    title = "{Wormhole solutions with scalar field and electric charge in modified gravity}",
    doi = "10.1088/1402-4896/abc985",
    journal = "Phys. Scripta",
    volume = "96",
    number = "1",
    pages = "015303",
    year = "2021"
}

@article{Tsukamoto:2012xs,
    author = "Tsukamoto, Naoki and Harada, Tomohiro and Yajima, Kohji",
    title = "{Can we distinguish between black holes and wormholes by their Einstein ring systems?}",
    eprint = "1207.0047",
    archivePrefix = "arXiv",
    primaryClass = "gr-qc",
    reportNumber = "RUP-12-5",
    doi = "10.1103/PhysRevD.86.104062",
    journal = "Phys. Rev. D",
    volume = "86",
    pages = "104062",
    year = "2012"
}

@article{Bambi:2013nla,
    author = "Bambi, Cosimo",
    title = "{Can the supermassive objects at the centers of galaxies be traversable wormholes? The first test of strong gravity for mm/sub-mm very long baseline interferometry facilities}",
    eprint = "1304.5691",
    archivePrefix = "arXiv",
    primaryClass = "gr-qc",
    doi = "10.1103/PhysRevD.87.107501",
    journal = "Phys. Rev. D",
    volume = "87",
    pages = "107501",
    year = "2013"
}

@article{Godani:2021aub,
    author = "Godani, Nisha and Samanta, Gauranga C.",
    title = "{Gravitational lensing effect in traversable wormholes}",
    eprint = "2105.08517",
    archivePrefix = "arXiv",
    primaryClass = "gr-qc",
    doi = "10.1016/j.aop.2021.168460",
    journal = "Annals Phys.",
    volume = "429",
    pages = "168460",
    year = "2021"
}

@article{Rosato:2025byu,
    author = "Rosato, Romeo Felice and Biswas, Shauvik and Chakraborty, Sumanta and Pani, Paolo",
    title = "{Greybody factors, reflectionless scattering modes, and echoes of ultracompact horizonless objects}",
    eprint = "2501.16433",
    archivePrefix = "arXiv",
    primaryClass = "gr-qc",
    doi = "10.1103/PhysRevD.111.084051",
    journal = "Phys. Rev. D",
    volume = "111",
    number = "8",
    pages = "084051",
    year = "2025"
}

@article{Rahaman:2025jhv,
    author = "Rahaman, Farook and Aziz, Abdul and Manna, Tuhina and Islam, Anikul and Pundeer, Naeem Ahmad and Islam, Sayeedul",
    title = "{Deflection of charged massive body around charged wormholes}",
    doi = "10.1007/s40042-025-01382-x",
    journal = "J. Korean Phys. Soc.",
    volume = "86",
    number = "12",
    pages = "1204--1224",
    year = "2025"
}

@article{Turimov:2025zxy,
    author = "Turimov, Bobur and Abdujabbarov, Ahmadjon and Ahmedov, Bobomurat and Stuchl{\'\i}k, Zden{\v{e}}k",
    title = "{Exact charged traversable wormhole solution}",
    doi = "10.1016/j.physletb.2025.139800",
    journal = "Phys. Lett. B",
    volume = "868",
    pages = "139800",
    year = "2025"
}

@article{Sarkar:2024bxk,
    author = "Sarkar, Nayan and Sarkar, Susmita and Bouzenada, Abdelmalek and Dutta, Abhisek and Sarkar, Moumita and Rahaman, Farook",
    title = "{Traversable wormholes with weak gravitational lensing effect in f(R,T) gravity}",
    doi = "10.1016/j.dark.2024.101439",
    journal = "Phys. Dark Univ.",
    volume = "44",
    pages = "101439",
    year = "2024"
}

@article{Dzhunushaliev:2025ngw,
    author = "Dzhunushaliev, Vladimir and Folomeev, Vladimir",
    title = "{Realistic classical charge from an asymmetric wormhole}",
    eprint = "2512.00958",
    archivePrefix = "arXiv",
    primaryClass = "gr-qc",
    doi = "10.1140/epjc/s10052-026-15596-1",
    journal = "Eur. Phys. J. C",
    volume = "86",
    number = "4",
    pages = "382",
    year = "2026"
}

@article{Blazquez-Salcedo:2025dit,
    author = "Bl{\'a}zquez-Salcedo, Jose Luis and Gonz{\'a}lez-Romero, Luis Manuel and Khoo, Fech Scen and Kunz, Jutta and Navarro Moreno, Pablo",
    title = "{Radial perturbations of charged wormholes}",
    eprint = "2510.11406",
    archivePrefix = "arXiv",
    primaryClass = "gr-qc",
    doi = "10.1016/j.physletb.2026.140335",
    journal = "Phys. Lett. B",
    volume = "875",
    pages = "140335",
    year = "2026"
}

@article{Gray:2021toe,
    author = "Gray, Finnian and Kubiznak, David",
    title = "{Slowly rotating black holes with exact Killing tensor symmetries}",
    eprint = "2110.14671",
    archivePrefix = "arXiv",
    primaryClass = "gr-qc",
    doi = "10.1103/PhysRevD.105.064017",
    journal = "Phys. Rev. D",
    volume = "105",
    number = "6",
    pages = "064017",
    year = "2022"
}

@article{Carter:1968rr,
    author = "Carter, Brandon",
    title = "{Global structure of the Kerr family of gravitational fields}",
    doi = "10.1103/PhysRev.174.1559",
    journal = "Phys. Rev.",
    volume = "174",
    pages = "1559--1571",
    year = "1968"
}

@article{Benenti:1979erw,
    author = "Benenti, S. and Francaviglia, M.",
    title = "{Remarks on certain separability structures and their applications to general relativity}",
    doi = "10.1007/bf00757025",
    journal = "Gen. Rel. Grav.",
    volume = "10",
    number = "1",
    pages = "79--92",
    year = "1979"
}

@article{Demianski:1980mgt,
    author = "Demianski, M. and Francaviglia, M.",
    title = "{Separability structures and Killing-Yano tensors in vacuum type-$D$ space-times without acceleration}",
    doi = "10.1007/bf00670402",
    journal = "Int. J. Theor. Phys.",
    volume = "19",
    number = "9",
    pages = "675--680",
    year = "1980"
}

@article{Kumar:2023wfp,
    author = "Kumar, Saurabh and Uniyal, Akhil and Chakrabarti, Sayan",
    title = "{Shadow and weak gravitational lensing of rotating traversable wormhole in nonhomogeneous plasma spacetime}",
    eprint = "2308.05545",
    archivePrefix = "arXiv",
    primaryClass = "gr-qc",
    doi = "10.1103/PhysRevD.109.104012",
    journal = "Phys. Rev. D",
    volume = "109",
    number = "10",
    pages = "104012",
    year = "2024"
}

@article{Frolov:2006pe,
    author = "Frolov, Valeri P. and Krtous, Pavel and Kubiznak, David",
    title = "{Separability of Hamilton-Jacobi and Klein-Gordon Equations in General Kerr-NUT-AdS Spacetimes}",
    eprint = "hep-th/0611245",
    archivePrefix = "arXiv",
    reportNumber = "ALBERTA-THY-15-06",
    doi = "10.1088/1126-6708/2007/02/005",
    journal = "JHEP",
    volume = "02",
    pages = "005",
    year = "2007"
}

@article{Kim:2025zyo,
    author = "Kim, Hyeong-Chan and Lee, Wonwoo",
    title = "{Charged wormholes in (anti-)de Sitter spacetime}",
    eprint = "2505.09981",
    archivePrefix = "arXiv",
    primaryClass = "gr-qc",
    reportNumber = "CQUeST-2025-0758",
    doi = "10.1016/j.physletb.2025.140145",
    journal = "Phys. Lett. B",
    volume = "873",
    pages = "140145",
    year = "2026"
}

@article{Kim:2001ri,
    author = "Kim, Sung-Won and Lee, Hyunjoo",
    title = "{Exact solutions of a charged wormhole}",
    eprint = "gr-qc/0102077",
    archivePrefix = "arXiv",
    doi = "10.1103/PhysRevD.63.064014",
    journal = "Phys. Rev. D",
    volume = "63",
    pages = "064014",
    year = "2001"
}

@article{Antoniou:2019awm,
    author = "Antoniou, Georgios and Bakopoulos, Athanasios and Kanti, Panagiota and Kleihaus, Burkhard and Kunz, Jutta",
    title = "{Novel Einstein{\textendash}scalar-Gauss-Bonnet wormholes without exotic matter}",
    eprint = "1904.13091",
    archivePrefix = "arXiv",
    primaryClass = "hep-th",
    doi = "10.1103/PhysRevD.101.024033",
    journal = "Phys. Rev. D",
    volume = "101",
    number = "2",
    pages = "024033",
    year = "2020"
}

@article{Churilova:2021tgn,
    author = "Churilova, M. S. and Konoplya, R. A. and Stuchlik, Z. and Zhidenko, A.",
    title = "{Wormholes without exotic matter: quasinormal modes, echoes and shadows}",
    eprint = "2107.05977",
    archivePrefix = "arXiv",
    primaryClass = "gr-qc",
    doi = "10.1088/1475-7516/2021/10/010",
    journal = "JCAP",
    volume = "10",
    pages = "010",
    year = "2021"
}

@article{PerezBergliaffa:2000ms,
    author = "Perez Bergliaffa, Santiago E. and Hibberd, K. E.",
    title = "{On the stress energy tensor of a rotating wormhole}",
    eprint = "gr-qc/0006041",
    archivePrefix = "arXiv",
    month = "6",
    year = "2000"
}

@article{Azad:2023iju,
    author = "Azad, Bahareh and Blazquez-Salcedo, Jose Luis and Khoo, Fech Scen and Kunz, Jutta",
    title = "{Are slowly rotating Ellis-Bronnikov wormholes stable?}",
    eprint = "2301.05243",
    archivePrefix = "arXiv",
    primaryClass = "gr-qc",
    doi = "10.1016/j.physletb.2023.138349",
    journal = "Phys. Lett. B",
    volume = "848",
    pages = "138349",
    year = "2024"
}

@article{Lima:2020auu,
    author = "Lima, Haroldo C. D. and Benone, Carolina L. and Crispino, Lu{\'\i}s C. B.",
    title = "{Scalar absorption: Black holes versus wormholes}",
    eprint = "2006.03967",
    archivePrefix = "arXiv",
    primaryClass = "gr-qc",
    doi = "10.1103/PhysRevD.101.124009",
    journal = "Phys. Rev. D",
    volume = "101",
    number = "12",
    pages = "124009",
    year = "2020"
}

@article{DeFalco:2023twb,
    author = "De Falco, Vittorio and Capozziello, Salvatore",
    title = "{Static and spherically symmetric wormholes in metric-affine theories of gravity}",
    eprint = "2308.05440",
    archivePrefix = "arXiv",
    primaryClass = "gr-qc",
    doi = "10.1103/PhysRevD.108.104030",
    journal = "Phys. Rev. D",
    volume = "108",
    number = "10",
    pages = "104030",
    year = "2023"
}

@article{LimaJunior:2022zvu,
    author = "Lima Junior, Haroldo C. D. and Benone, Carolina L. and Crispino, Lu{\'\i}s C. B.",
    title = "{Scalar scattering by black holes and wormholes}",
    eprint = "2211.09886",
    archivePrefix = "arXiv",
    primaryClass = "gr-qc",
    doi = "10.1140/epjc/s10052-022-10576-7",
    journal = "Eur. Phys. J. C",
    volume = "82",
    number = "7",
    pages = "638",
    year = "2022"
}

@article{Gyulchev:2018fmd,
    author = "Gyulchev, Galin and Nedkova, Petya and Tinchev, Vassil and Yazadjiev, Stoytcho",
    title = "{On the shadow of rotating traversable wormholes}",
    eprint = "1805.11591",
    archivePrefix = "arXiv",
    primaryClass = "gr-qc",
    doi = "10.1140/epjc/s10052-018-6012-9",
    journal = "Eur. Phys. J. C",
    volume = "78",
    number = "7",
    pages = "544",
    year = "2018"
}

@article{Konoplya:2010kv,
    author = "Konoplya, R. A. and Zhidenko, A.",
    title = "{Passage of radiation through wormholes of arbitrary shape}",
    eprint = "1004.1284",
    archivePrefix = "arXiv",
    primaryClass = "hep-th",
    doi = "10.1103/PhysRevD.81.124036",
    journal = "Phys. Rev. D",
    volume = "81",
    pages = "124036",
    year = "2010"
}

@article{Azreg-Ainou:2014dwa,
    author = {Azreg-A{\"\i}nou, Mustapha},
    title = "{Confined-exotic-matter wormholes with no gluing effects{\textemdash}Imaging supermassive wormholes and black holes}",
    eprint = "1412.8282",
    archivePrefix = "arXiv",
    primaryClass = "gr-qc",
    doi = "10.1088/1475-7516/2015/07/037",
    journal = "JCAP",
    volume = "07",
    pages = "037",
    year = "2015"
}

@article{Abdujabbarov:2016efm,
    author = "Abdujabbarov, Ahmadjon and Juraev, Bakhtinur and Ahmedov, Bobomurat and Stuchl{\'\i}k, Zden{\v{e}}k",
    title = "{Shadow of rotating wormhole in plasma environment}",
    doi = "10.1007/s10509-016-2818-9",
    journal = "Astrophys. Space Sci.",
    volume = "361",
    number = "7",
    pages = "226",
    year = "2016"
}

@article{Dai:2019mse,
    author = "Dai, De-Chang and Stojkovic, Dejan",
    title = "{Observing a Wormhole}",
    eprint = "1910.00429",
    archivePrefix = "arXiv",
    primaryClass = "gr-qc",
    doi = "10.1103/PhysRevD.100.083513",
    journal = "Phys. Rev. D",
    volume = "100",
    number = "8",
    pages = "083513",
    year = "2019"
}

@article{Simonetti:2020ivl,
    author = "Simonetti, John H. and Kavic, Michael J. and Minic, Djordje and Stojkovic, Dejan and Dai, De-Chang",
    title = "{Sensitive searches for wormholes}",
    eprint = "2007.12184",
    archivePrefix = "arXiv",
    primaryClass = "gr-qc",
    doi = "10.1103/PhysRevD.104.L081502",
    journal = "Phys. Rev. D",
    volume = "104",
    number = "8",
    pages = "L081502",
    year = "2021"
}

@article{Teo:1998dp,
    author = "Teo, Edward",
    title = "{Rotating traversable wormholes}",
    eprint = "gr-qc/9803098",
    archivePrefix = "arXiv",
    reportNumber = "DAMTP-R-98-17",
    doi = "10.1103/PhysRevD.58.024014",
    journal = "Phys. Rev. D",
    volume = "58",
    pages = "024014",
    year = "1998"
}

@book{abramowitz1964handbook,
  title        = "Handbook of Mathematical Functions with Formulas, Graphs, and Mathematical Tables",
  author       = "Milton Abramowitz and Irene A. Stegun",
  year         = "1964",
  publisher    = "U.S. Government Printing Office",
  address      = "Washington, D.C.",
  series       = "National Bureau of Standards Applied Mathematics Series",
  volume       = "55",
  note         = "Reprinted by Dover Publications, 1965",
  isbn         = "9780486612720"
}

@article{Benone:2019all,
    author = "Benone, Carolina L. and Crispino, Lu{\'\i}s C. B.",
    title = "{Massive and charged scalar field in Kerr-Newman spacetime: Absorption and superradiance}",
    eprint = "1901.05592",
    archivePrefix = "arXiv",
    primaryClass = "gr-qc",
    doi = "10.1103/PhysRevD.99.044009",
    journal = "Phys. Rev. D",
    volume = "99",
    number = "4",
    pages = "044009",
    year = "2019"
}

@misc{sakurai1986modern,
  title={Modern quantum mechanics},
  author={Sakurai, Jun John and Fu Tuan, San and Newton, Roger G},
  year={1986},
  publisher={American Institute of Physics}
}

@article{Macedo:2018yoi,
    author = "Macedo, Caio F. B. and Stratton, Tom and Dolan, Sam and Crispino, Lu{\'\i}s C. B.",
    title = "{Spectral lines of extreme compact objects}",
    eprint = "1807.04762",
    archivePrefix = "arXiv",
    primaryClass = "gr-qc",
    doi = "10.1103/PhysRevD.98.104034",
    journal = "Phys. Rev. D",
    volume = "98",
    number = "10",
    pages = "104034",
    year = "2018"
}

@article{Macedo:2013afa,
    author = "Macedo, Caio F. B. and Leite, Lu{\'\i}z C. S. and Oliveira, Ednilton S. and Dolan, Sam R. and Crispino, Lu{\'\i}s C. B.",
    title = "{Absorption of planar massless scalar waves by Kerr black holes}",
    eprint = "1308.0018",
    archivePrefix = "arXiv",
    primaryClass = "gr-qc",
    doi = "10.1103/PhysRevD.88.064033",
    journal = "Phys. Rev. D",
    volume = "88",
    number = "6",
    pages = "064033",
    year = "2013"
}

@article{Nedkova:2013msa,
    author = "Nedkova, Petya G. and Tinchev, Vassil K. and Yazadjiev, Stoytcho S.",
    title = "{Shadow of a rotating traversable wormhole}",
    eprint = "1307.7647",
    archivePrefix = "arXiv",
    primaryClass = "gr-qc",
    doi = "10.1103/PhysRevD.88.124019",
    journal = "Phys. Rev. D",
    volume = "88",
    number = "12",
    pages = "124019",
    year = "2013"
}

@article{Hochberg:1998ii,
    author = "Hochberg, David and Visser, Matt",
    title = "{The Null energy condition in dynamic wormholes}",
    eprint = "gr-qc/9802048",
    archivePrefix = "arXiv",
    reportNumber = "LAEFF-98-02",
    doi = "10.1103/PhysRevLett.81.746",
    journal = "Phys. Rev. Lett.",
    volume = "81",
    pages = "746--749",
    year = "1998"
}

@article{LIGOScientific:2021sio,
    author = "Abbott, R. and others",
    collaboration = "LIGO Scientific, VIRGO, KAGRA",
    title = "{Tests of General Relativity with GWTC-3}",
    eprint = "2112.06861",
    archivePrefix = "arXiv",
    primaryClass = "gr-qc",
    reportNumber = "LIGO-P2100275",
    doi = "10.1103/PhysRevD.112.084080",
    journal = "Phys. Rev. D",
    volume = "112",
    number = "8",
    pages = "084080",
    year = "2025"
}

@article{Vagnozzi:2022moj,
    author = "Vagnozzi, Sunny and others",
    title = "{Horizon-scale tests of gravity theories and fundamental physics from the Event Horizon Telescope image of Sagittarius A}",
    eprint = "2205.07787",
    archivePrefix = "arXiv",
    primaryClass = "gr-qc",
    reportNumber = "UCI-HEP-TR-2022-07",
    doi = "10.1088/1361-6382/acd97b",
    journal = "Class. Quant. Grav.",
    volume = "40",
    number = "16",
    pages = "165007",
    year = "2023"
}

@article{Riley:2026avn,
    author = "Riley, Jeff",
    title = "{Electromagnetic, gravitational wave, and static gravitational transmission through throat spacetimes: a constraint-wave asymmetry}",
    eprint = "2604.14238",
    archivePrefix = "arXiv",
    primaryClass = "gr-qc",
    month = "4",
    year = "2026"
}

@article{Kar:1995jz,
    author = "Kar, Sayan and Minwalla, Shiraz and Mishra, D. and Sahdev, D.",
    title = "{Resonances in the transmission of massless scalar waves in a class of wormholes}",
    doi = "10.1103/PhysRevD.51.1632",
    journal = "Phys. Rev. D",
    volume = "51",
    pages = "1632--1638",
    year = "1995"
}

@inproceedings{Lobo:2020xvs,
    author = "Lobo, Francisco S. N. and Rubiera-Garcia, Diego",
    title = "{Wormholes, energy conditions and time machines}",
    booktitle = "{15th Marcel Grossmann Meeting on Recent Developments in Theoretical and Experimental General Relativity, Astrophysics, and Relativistic Field Theories}",
    eprint = "2008.09902",
    archivePrefix = "arXiv",
    primaryClass = "gr-qc",
    doi = "10.1142/9789811258251_0074",
    month = "8",
    year = "2020"
}

@article{Hoffmann:2018oml,
    author = "Hoffmann, Christian and Ioannidou, Theodora and Kahlen, Sarah and Kleihaus, Burkhard and Kunz, Jutta",
    title = "{Symmetric and Asymmetric Wormholes Immersed In Rotating Matter}",
    eprint = "1803.11044",
    archivePrefix = "arXiv",
    primaryClass = "gr-qc",
    doi = "10.1103/PhysRevD.97.124019",
    journal = "Phys. Rev. D",
    volume = "97",
    number = "12",
    pages = "124019",
    year = "2018"
}

@article{Forghani:2018gza,
    author = "Forghani, S. Danial and Habib Mazharimousavi, S. and Halilsoy, Mustafa",
    title = "{Asymmetric Thin-Shell Wormholes}",
    eprint = "1801.05516",
    archivePrefix = "arXiv",
    primaryClass = "gr-qc",
    doi = "10.1140/epjc/s10052-018-5776-2",
    journal = "Eur. Phys. J. C",
    volume = "78",
    number = "6",
    pages = "469",
    year = "2018"
}

@article{Forghani:2018fks,
    author = "Forghani, S. Danial and Habib Mazharimousavi, S. and Halilsoy, M.",
    title = "{Cylindrical asymmetric thin-shell wormholes}",
    eprint = "1807.05080",
    archivePrefix = "arXiv",
    primaryClass = "gr-qc",
    doi = "10.1088/1475-7516/2019/10/067",
    journal = "JCAP",
    volume = "10",
    pages = "067",
    year = "2019"
}

@article{Uemichi:2026dzb,
    author = "Uemichi, Keiya and Koga, Yasutaka and Saito, Daiki and Yoo, Chul-Moon and Yoshida, Daisuke",
    title = "{Rotating wormholes in five dimensions with equal angular momenta: large asymmetry regime}",
    eprint = "2603.12748",
    archivePrefix = "arXiv",
    primaryClass = "gr-qc",
    reportNumber = "NU-QG-14, KUNS-3095",
    month = "3",
    year = "2026"
}

@article{Kashargin:2008pk,
    author = "Kashargin, P. E. and Sushkov, S. V.",
    title = "{Slowly rotating scalar field wormholes: The Second order approximation}",
    eprint = "0809.1923",
    archivePrefix = "arXiv",
    primaryClass = "gr-qc",
    doi = "10.1103/PhysRevD.78.064071",
    journal = "Phys. Rev. D",
    volume = "78",
    pages = "064071",
    year = "2008"
}

@article{Harko:2009xf,
    author = "Harko, Tiberiu and Kovacs, Zoltan and Lobo, Francisco S. N.",
    title = "{Thin accretion disks in stationary axisymmetric wormhole spacetimes}",
    eprint = "0901.3926",
    archivePrefix = "arXiv",
    primaryClass = "gr-qc",
    doi = "10.1103/PhysRevD.79.064001",
    journal = "Phys. Rev. D",
    volume = "79",
    pages = "064001",
    year = "2009"
}

@article{Karimov:2019qfw,
    author = "Karimov, R. Kh. and Izmailov, R. N. and Nandi, K. K.",
    title = "{Accretion disk around the rotating Damour{\textendash}Solodukhin wormhole}",
    eprint = "1901.05762",
    archivePrefix = "arXiv",
    primaryClass = "gr-qc",
    doi = "10.1140/epjc/s10052-019-7488-7",
    journal = "Eur. Phys. J. C",
    volume = "79",
    number = "11",
    pages = "952",
    year = "2019"
}

@article{Paul:2019trt,
    author = "Paul, Suvankar and Shaikh, Rajibul and Banerjee, Pritam and Sarkar, Tapobrata",
    title = "{Observational signatures of wormholes with thin accretion disks}",
    eprint = "1911.05525",
    archivePrefix = "arXiv",
    primaryClass = "gr-qc",
    doi = "10.1088/1475-7516/2020/03/055",
    journal = "JCAP",
    volume = "03",
    pages = "055",
    year = "2020"
}

@article{Cisterna:2023uqf,
    author = {Cisterna, Adolfo and M{\"u}ller, Keanu and Pallikaris, Konstantinos and Vigan{\`o}, Adriano},
    title = "{Exact rotating wormholes via Ehlers transformations}",
    eprint = "2306.14541",
    archivePrefix = "arXiv",
    primaryClass = "gr-qc",
    doi = "10.1103/PhysRevD.108.024066",
    journal = "Phys. Rev. D",
    volume = "108",
    number = "2",
    pages = "024066",
    year = "2023"
}

@article{Volkov:2021blw,
    author = "Volkov, Mikhail S.",
    title = "{Stationary generalizations for the Bronnikov-Ellis wormhole and for the vacuum ring wormhole}",
    eprint = "2109.14496",
    archivePrefix = "arXiv",
    primaryClass = "gr-qc",
    doi = "10.1103/PhysRevD.104.124064",
    journal = "Phys. Rev. D",
    volume = "104",
    number = "12",
    pages = "124064",
    year = "2021"
}

@article{kashargin2008slowly,
  title={Slowly rotating scalar field wormholes: The second order approximation},
  author={Kashargin, PE and Sushkov, SV},
  journal={Physical Review D—Particles, Fields, Gravitation, and Cosmology},
  volume={78},
  number={6},
  pages={064071},
  year={2008},
  publisher={APS}
}

@article{Kleihaus:2014dla,
    author = "Kleihaus, Burkhard and Kunz, Jutta",
    title = "{Rotating Ellis Wormholes in Four Dimensions}",
    eprint = "1409.1503",
    archivePrefix = "arXiv",
    primaryClass = "gr-qc",
    doi = "10.1103/PhysRevD.90.121503",
    journal = "Phys. Rev. D",
    volume = "90",
    pages = "121503",
    year = "2014"
}

@article{Chew:2016epf,
    author = "Chew, Xiao Yan and Kleihaus, Burkhard and Kunz, Jutta",
    title = "{Geometry of Spinning Ellis Wormholes}",
    eprint = "1608.05253",
    archivePrefix = "arXiv",
    primaryClass = "gr-qc",
    doi = "10.1103/PhysRevD.94.104031",
    journal = "Phys. Rev. D",
    volume = "94",
    number = "10",
    pages = "104031",
    year = "2016"
}

@article{Unruh:1976fm,
    author = "Unruh, W. G.",
    title = "{Absorption Cross-Section of Small Black Holes}",
    doi = "10.1103/PhysRevD.14.3251",
    journal = "Phys. Rev. D",
    volume = "14",
    pages = "3251--3259",
    year = "1976"
}

@article{Cardoso:2019dte,
    author = "Cardoso, Vitor and Vicente, Rodrigo",
    title = "{Moving black holes: energy extraction, absorption cross-section and the ring of fire}",
    eprint = "1906.10140",
    archivePrefix = "arXiv",
    primaryClass = "gr-qc",
    doi = "10.1103/PhysRevD.100.084001",
    journal = "Phys. Rev. D",
    volume = "100",
    number = "8",
    pages = "084001",
    year = "2019"
}

@article{Brito:2015oca,
    author = "Brito, Richard and Cardoso, Vitor and Pani, Paolo",
    title = "{Superradiance}: {New Frontiers in Black Hole
Physics}",
    eprint = "1501.06570",
    archivePrefix = "arXiv",
    primaryClass = "gr-qc",
    doi = "10.1007/978-3-319-19000-6",
    journal = "Lect. Notes Phys.",
    volume = "906",
    pages = "pp.1--237",
    year = "2015"
}

@article{DeFalco:2023kqy,
    author = "De Falco, Vittorio",
    title = "{Epicyclic frequencies in the equatorial plane around stationary and axially symmetric wormhole geometries}",
    eprint = "2307.03151",
    archivePrefix = "arXiv",
    primaryClass = "gr-qc",
    doi = "10.1103/PhysRevD.108.024051",
    journal = "Phys. Rev. D",
    volume = "108",
    number = "2",
    pages = "024051",
    year = "2023"
}

@article{DeSimone:2025sgu,
    author = "De Simone, Ciro and De Falco, Vittorio and Capozziello, Salvatore",
    title = "{Can wormholes mirror the quasinormal mode spectrum of Schwarzschild black holes?}",
    eprint = "2502.12646",
    archivePrefix = "arXiv",
    primaryClass = "gr-qc",
    doi = "10.1103/PhysRevD.111.064021",
    journal = "Phys. Rev. D",
    volume = "111",
    number = "6",
    pages = "064021",
    year = "2025"
}

@article{Biswas:2023ofz,
    author = "Biswas, Shauvik and Singha, Chiranjeeb and Chakraborty, Sumanta",
    title = "{Galactic wormholes: Geometry, stability, and echoes}",
    eprint = "2307.04836",
    archivePrefix = "arXiv",
    primaryClass = "gr-qc",
    doi = "10.1103/PhysRevD.109.064043",
    journal = "Phys. Rev. D",
    volume = "109",
    number = "6",
    pages = "064043",
    year = "2024"
}

@article{Rosato:2025lxb,
    author = "Rosato, Romeo Felice and Biswas, Shauvik and Chakraborty, Sumanta and Pani, Paolo",
    title = "{Excitation factors for horizonless compact objects: Long-lived modes, echoes, and greybody factors}",
    eprint = "2511.08692",
    archivePrefix = "arXiv",
    primaryClass = "gr-qc",
    doi = "10.1103/v7pv-y3qy",
    journal = "Phys. Rev. D",
    volume = "113",
    number = "8",
    pages = "084002",
    year = "2026"
}

@article{Cardoso:2016rao,
    author = "Cardoso, Vitor and Franzin, Edgardo and Pani, Paolo",
    title = "{Is the gravitational-wave ringdown a probe of the event horizon?}",
    eprint = "1602.07309",
    archivePrefix = "arXiv",
    primaryClass = "gr-qc",
    doi = "10.1103/PhysRevLett.116.171101",
    journal = "Phys. Rev. Lett.",
    volume = "116",
    number = "17",
    pages = "171101",
    year = "2016",
    note = "[Erratum: Phys.Rev.Lett. 117, 089902 (2016)]"
}

@article{Konoplya:2018ala,
    author = "Konoplya, R. A.",
    title = "{How to tell the shape of a wormhole by its quasinormal modes}",
    eprint = "1805.04718",
    archivePrefix = "arXiv",
    primaryClass = "gr-qc",
    doi = "10.1016/j.physletb.2018.07.025",
    journal = "Phys. Lett. B",
    volume = "784",
    pages = "43--49",
    year = "2018"
}

@article{Konoplya:2025hgp,
    author = "Konoplya, R. A. and Stashko, O. S.",
    title = "{Transition from regular black holes to wormholes in covariant effective quantum gravity: Scattering, quasinormal modes, and Hawking radiation}",
    eprint = "2502.05689",
    archivePrefix = "arXiv",
    primaryClass = "gr-qc",
    doi = "10.1103/PhysRevD.111.084031",
    journal = "Phys. Rev. D",
    volume = "111",
    number = "8",
    pages = "084031",
    year = "2025"
}

@article{Chandrasekhar:1975zza,
    author = "Chandrasekhar, S. and Detweiler, Steven L.",
    title = "{The quasi-normal modes of the Schwarzschild black hole}",
    doi = "10.1098/rspa.1975.0112",
    journal = "Proc. Roy. Soc. Lond. A",
    volume = "344",
    pages = "441--452",
    year = "1975"
}

@article{Cardoso:2007az,
    author = "Cardoso, Vitor and Pani, Paolo and Cadoni, Mariano and Cavaglia, Marco",
    title = "{Ergoregion instability of ultracompact astrophysical objects}",
    eprint = "0709.0532",
    archivePrefix = "arXiv",
    primaryClass = "gr-qc",
    doi = "10.1103/PhysRevD.77.124044",
    journal = "Phys. Rev. D",
    volume = "77",
    pages = "124044",
    year = "2008"
}

@article{Maggio:2017ivp,
    author = "Maggio, Elisa and Pani, Paolo and Ferrari, Valeria",
    title = "{Exotic Compact Objects and How to Quench their Ergoregion Instability}",
    eprint = "1703.03696",
    archivePrefix = "arXiv",
    primaryClass = "gr-qc",
    doi = "10.1103/PhysRevD.96.104047",
    journal = "Phys. Rev. D",
    volume = "96",
    number = "10",
    pages = "104047",
    year = "2017"
}

@article{Friedman:1978wla,
    author = "Friedman, John L.",
    title = "{Generic instability of rotating relativistic stars}",
    doi = "10.1007/BF01202527",
    journal = "Commun. Math. Phys.",
    volume = "62",
    number = "3",
    pages = "247--278",
    year = "1978"
}

@article{Siemonsen:2025wib,
    author = "Siemonsen, Nils",
    title = "{Ergoregion instability in bosonic stars: Scalar mode structure, universality, and weakly nonlinear effects}",
    eprint = "2510.07468",
    archivePrefix = "arXiv",
    primaryClass = "gr-qc",
    doi = "10.1103/sbtt-jnjg",
    journal = "Phys. Rev. D",
    volume = "112",
    number = "12",
    pages = "124031",
    year = "2025"
}

@article{Benone:2014qaa,
    author = "Benone, Carolina L. and de Oliveira, Ednilton S. and Dolan, Sam R. and Crispino, Lu{\'\i}s C. B.",
    title = "{Absorption of a massive scalar field by a charged black hole}",
    eprint = "1404.0687",
    archivePrefix = "arXiv",
    primaryClass = "gr-qc",
    doi = "10.1103/PhysRevD.89.104053",
    journal = "Phys. Rev. D",
    volume = "89",
    number = "10",
    pages = "104053",
    year = "2014"
}

@article{Karmakar:2026yqg,
    author = "Karmakar, Rajesh and Ge, Xian-Hui",
    title = "{Analytical solution of traversable wormholes in the presence of positive cosmological constant}",
    eprint = "2605.08205",
    archivePrefix = "arXiv",
    primaryClass = "gr-qc",
    doi = "10.1088/1475-7516/2026/09/060",
    journal = "JCAP",
    volume = "09",
    pages = "060",
    year = "2026"
}

@article{Karmakar:2024uzt,
    author = "Karmakar, Rajesh",
    title = "{Superradiant scattering of massive scalar field due to magnetically charged rotating black hole}",
    eprint = "2408.11410",
    archivePrefix = "arXiv",
    primaryClass = "gr-qc",
    doi = "10.1016/j.physletb.2025.139392",
    journal = "Phys. Lett. B",
    volume = "863",
    pages = "139392",
    year = "2025"
}

@article{Franzin:2022iai,
    author = "Franzin, Edgardo and Liberati, Stefano and Mazza, Jacopo and Dey, Ramit and Chakraborty, Sumanta",
    title = "{Scalar perturbations around rotating regular black holes and wormholes: Quasinormal modes, ergoregion instability, and superradiance}",
    eprint = "2201.01650",
    archivePrefix = "arXiv",
    primaryClass = "gr-qc",
    doi = "10.1103/PhysRevD.105.124051",
    journal = "Phys. Rev. D",
    volume = "105",
    number = "12",
    pages = "124051",
    year = "2022"
}

@article{Mazza:2021rgq,
    author = "Mazza, Jacopo and Franzin, Edgardo and Liberati, Stefano",
    title = "{A novel family of rotating black hole mimickers}",
    eprint = "2102.01105",
    archivePrefix = "arXiv",
    primaryClass = "gr-qc",
    doi = "10.1088/1475-7516/2021/04/082",
    journal = "JCAP",
    volume = "04",
    pages = "082",
    year = "2021"
}
%%%%%%%%%%%%%%%%%%%%%%%%%%%%%%%%%%%%
%%%%%%%%%%%%%%%%%%%%%%%%%%%%%%%%%%%%
\end{document}